\documentclass[aps,prb,amsmath,amssymb,reprint,groupedaddress]{revtex4-2}

\usepackage{graphicx}
\usepackage{dcolumn}
\usepackage{bm}
\usepackage{hyperref}
\usepackage{siunitx}
\usepackage[english]{babel}

\begin{document}

\newcommand{\AEphonon}{{\mathrm{A}_{1}}^{\prime}/\mathrm{E}^{\prime}}


\title{Resonant Raman signatures of bright and momentum-dark exciton coupled by intervalley phonon scattering in monolayer WSe\textsubscript{2}}


\author{Hendrik Lambers}
\email[]{hendriklambers@uni-muenster.de}
\altaffiliation{}
\affiliation{Institute of Physics, University of Münster, Münster, Germany}
\affiliation{Center for Soft Nanoscience (SoN), University of Münster, Münster,
Germany}

\author{Nihit Saigal}
\affiliation{Institute of Physics, University of Münster, Münster, Germany}
\affiliation{Center for Soft Nanoscience (SoN), University of Münster, Münster,
Germany}

\author{Lara Blinov}
\affiliation{Institute of Physics, University of Münster, Münster, Germany}
\affiliation{Center for Soft Nanoscience (SoN), University of Münster, Münster,
Germany}

\author{Jonas Kiemle}
\affiliation{Walter Schottky Institut and Physics Department, Technical University of Munich, Garching, Germany}
\affiliation{MCQST, München, Germany}

\author{Alexander W. Holleitner}
\affiliation{Walter Schottky Institut and Physics Department, Technical University of Munich, Garching, Germany}
\affiliation{MCQST, München, Germany}

\author{Ursula Wurstbauer}
\email{wurstbauer@uni-muenster.de}
\affiliation{Institute of Physics, University of Münster, Münster, Germany}
\affiliation{Center for Soft Nanoscience (SoN), University of Münster, Münster,
Germany}


\date{\today}

\begin{abstract}

Exciton-phonon coupling in atomically thin transition metal dichalcogenides governs key processes such as exciton thermalization and intervalley scattering and remains challenging to access directly by optical spectroscopy. Here, we employ resonance Raman spectroscopy at cryogenic temperatures to probe exciton-phonon coupling in hBN-encapsulated WSe\textsubscript{2} monolayers. Tuning the excitation laser across the bright exciton $\mathrm{X_{KK}}$ resonance, we observe rich Raman spectra and focus on the resonance profile of the degenerate $\AEphonon$ optical phonon mode. The profile exhibits two asymmetric resonance peaks whose energetic separation significantly exceeds the phonon energy — a feature that cannot be explained by first-order Raman scattering alone. We demonstrate that this discrepancy is resolved by including third-order Raman scattering, in which intervalley scattering enabled by a finite-momentum phonon couples the bright exciton $\mathrm{X_{KK}}$ to a momentum-dark exciton $\mathrm{X_{d}}$. Fitting the experimental resonance profiles of two independent samples with a model comprising coherent first- and third-order scattering yields consistent exciton energies and linewidths, with a momentum-dark exciton $\mathrm{X_{d}}$ approximately $\SIrange{45}{55}{\milli\electronvolt}$ below the bright exciton $\mathrm{X_{KK}}$. The results indicate efficient bright-to-dark exciton coupling. Our findings provide a microscopic framework for understanding the anomalously bright emission spectra of WSe\textsubscript{2} monolayers despite its spin-forbidden lowest exciton transition and highlight the role of momentum-dark excitons in resonant light-matter interaction.
\end{abstract}

\keywords{WSe\textsubscript{2}, resonance Raman spectroscopy, exciton intervalley scattering}

\maketitle

\section{\label{sec:Introduction}Introduction}


Atomically thin transition metal dichalcogenides (TMDCs) have emerged as a compelling platform for investigating light-matter coupling in the two dimensional limit \cite{wang_electronics_2012, mueller_exciton_2018}. In the monolayer group-VI TMDCs such as MoSe\textsubscript{2} and WSe\textsubscript{2}, the transition from an indirect to a direct bandgap gives rise to excitons with exceptionally large binding energies, that dominate the light-matter interaction \cite{mak_atomically_2010, splendiani_emerging_2010, chernikov_exciton_2014}. While MoSe\textsubscript{2} hosts an optically bright exciton as its lowest-energy excitonic state, the excitonic landscape of WSe\textsubscript{2} is considerably richer: multiple momentum-dark and spin-dark exciton-states lie energetically below the optically bright exciton $\mathrm{X_{KK}}$ \cite{deilmann_dark_2017, malic_dark_2018}, as illustrated schematically in Fig. \ref{fig:Figure_A}(a). This energetic ordering renders WSe\textsubscript{2} effectively a momentum-indirect semiconductor and makes it an exceptional platform for studying the interplay between bright and dark exciton states. 

The presence of multiple low-lying dark exciton states has profound implications for the optical and electronic dynamics of WSe\textsubscript{2}. Exciton thermalization, population redistribution, and intervalley scattering are all governed by exciton-phonon coupling involving finite-momentum phonons that connect the bright exciton $\mathrm{X_{KK}}$ to momentum-dark excitons \cite{selig_dark_2018, niehues_strain_2018, bange_ultrafast_2023}. These processes ultimately determine the emission efficiency and the anomalously bright photoluminescence (PL) observed in WSe\textsubscript{2} monolayers despite the spin-forbidden character of the lowest exciton transition \cite{tonndorf_photoluminescence_2013, arora_excitonic_2015, lindlau_role_2018, brem_phonon-assisted_2020}. A quantitative understanding of the relevant coupling strengths and scattering channels is therefore essential — yet experimentally challenging to access directly.

Standard optical spectroscopy techniques provide only indirect access to exciton-phonon coupling mechanisms. Widely used PL primarily probes radiative recombination of thermalized exciton populations, not providing distinct information over the relaxation pathways that precedes the emission event \cite{selig_dark_2018}. Differential reflection spectroscopy is sensitive to  optical absorption at bright exciton transitions, but yields no direct information on the phonon modes or scattering channels after the absorption process \cite{qiu_optical_2013}. In both methods, the microscopic exciton-phonon coupling matrix elements and the individual scattering channels via dark exciton states remain largely concealed.

Resonance Raman spectroscopy (RRS) provides a powerful access to the exciton-phonon coupling. When the laser excitation energy is tuned into resonance with a bright exciton, Raman scattering probes the phonon modes via intermediate exciton states — directly sensitive to the exciton-phonon coupling without being masked by thermalization dynamics \cite{del_corro_excited_2014, miller_tuning_2019, jadczak_resonant_2025}. RRS has been employed extensively in the context of graphene \cite{saito_probing_2001, cancado_stokes_2002, malard_raman_2009} and carbon nanotubes \cite{duque_violation_2011, haroz_asymmetric_2015, gordeev_asymmetry_2019} to reveal higher-order scattering mechanisms, and has been applied to TMDC monolayers to probe zone-center and zone-edge phonon modes \cite{scheuschner_resonant_2012, del_corro_excited_2014, golasa_multiphonon_2014, soubelet_resonance_2016, carvalho_intervalley_2017, mcdonnell_probing_2018, mcdonnell_observation_2020, zinkiewicz_raman_2024}. However, a systematic analysis of the role of momentum-dark excitons as intermediate scattering states from the resonance profiles of individual phonon modes in WSe\textsubscript{2} has not been established.
 
In this work, we address this gap by performing systematic RRS on hexagonal boron nitride (hBN)-encapsulated WSe\textsubscript{2} monolayers at cryogenic temperatures by continuously and precisely tuning the excitation energy across the resonance of the bright exciton $\mathrm{X_{KK}}$. Focusing on the resonance profile of the Raman-active degenerate $\AEphonon$ optical phonon mode, we identify a characteristic two-peak structure whose peak separation significantly exceeds the $\AEphonon$ phonon energy — a discrepancy that a standard first-order Raman scattering model fails to reproduce. We demonstrate that this discrepancy is quantitatively resolved by extending the model to include coherent third-order Raman scattering channels, in which finite-momentum phonon scattering facilitates intervalley coupling between the bright exciton $\mathrm{X_{KK}}$ and a momentum-dark exciton $\mathrm{X_d}$ \cite{niehues_strain_2018, bange_ultrafast_2023}. The energy detuning between the exciton species is sample-specific and known to be sensitive to strain, doping, and the local dielectric environment \cite{chernikov_electrical_2015, schmidt_reversible_2016, raja_coulomb_2017}. Our analysis therefore provides a means to extract their energy detuning. The efficient phonon-mediated bright-to-dark exciton coupling inferred from the analysis further motivates an interpretation in terms of linear combinations of exciton states, which naturally suggests hybridized bright-dark exciton states \cite{kumar_strain_2024,kumar_strain_2025}. Taken together, our results demonstrate resonance profile analysis as a sensitive probe of momentum-dark exciton energies and exciton-phonon coupling in WSe\textsubscript{2}. This offers a microscopic perspective on the material's anomalous optical properties.

The paper is organized as follows. Section \ref{sec:methods} recaps the experimental methods. Section \ref{sec:experimental results} presents the experimental results, introduces the first-order Raman scattering model, and identifies its failure to reproduce the observed resonance profiles. Section \ref{sec:intervalley_scattering_model} develops the extended third-order intervalley scattering model and discusses the results. Section \ref{sec:conclusion} concludes with a summary and outlook.

\section{\label{sec:methods}Methods}

hBN encapsulated WSe\textsubscript{2} monolayer samples were prepared by micromechanical cleavage and viscoelastic dry transfer onto Si/SiO\textsubscript{2} substrates. Details on the sample fabrication are given in the Supplemental Information. After initial characterization by room temperature Raman and PL measurements, the samples were cooled in a closed cycle $^3$He/$^4$He dilution refrigerator  equipped with optical windows for direct optical access. All measurements were conducted at $\SI{4}{\kelvin}$ or below. The continuous-wave output of an ultra-stable, precisely tunable ring-cavity Ti:sapphire laser was focused onto the sample using a low temperature objective (NA = 0.82), resulting in a spot size of less than $\SI{2}{\micro\meter}$. Sample positioning and focusing was controlled by piezo actuators. Scattered light was spectrally resolved using a triple spectrometer operated in subtractive mode and detected with a liquid-nitrogen-cooled CCD camera. The detected photon counts were normalized to the integration time and incident photon number per second. For RRS measurements the excitation wavelength was tuned in the range from $\SI{1.68}{\electronvolt}$ to $\SI{1.82}{\electronvolt}$ ($\SI{739}{\nano\meter}$ to $\SI{680}{\nano\meter}$ with a minimal step size of $\SI{1}{\nano\meter}$). This range includes the resonance with the bright exciton $\mathrm{X_{KK}}$, reported in the literature at cryogenic temperatures at approximately $\SI{1.735}{\electronvolt}$\cite{jones_optical_2013, arora_excitonic_2015}.

\section{\label{sec:experimental results}Experimental results and 1\textsuperscript{st} order scattering model}

\begin{figure*}
    \includegraphics[]{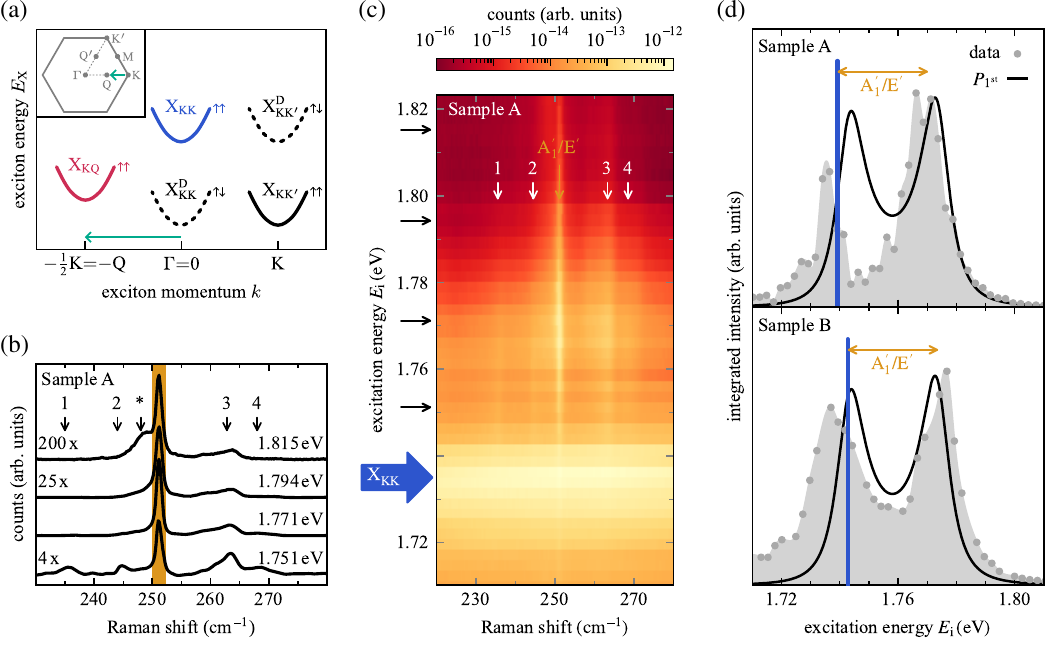}
    \caption{\label{fig:Figure_A}
    RRS measurement on a WSe\textsubscript{2} monolayer. 
    (a) Schematic exciton dispersion at high symmetry points. Subscripts denote hole (first index) and electron (second index) momentum. The bright exciton $\mathrm{X_{KK}}$ (blue) has zero center-of-mass momentum $k=0$ and parallel spins. The momentum-dark excitons $\mathrm{X_{KQ}}$ (red) and $\mathrm{X_{KK'}}$ (black) carry finite-momentum and lie at lower energies. Spin dark excitons $\mathrm{X_{KK}^D}$ and $\mathrm{X_{KK'}^D}$ (dashed) are energetically degenerate with the corresponding spin-flipped states. The momentum mismatch between $\mathrm{X_{KK}}$ and $\mathrm{X_{KQ}}$ is shown as a green arrow in the inset showing the first Brillouin zone (grey). 
    (b) Raman spectra recorded at selected excitation energies ranging from $\SI{1.751}{\electronvolt}$ to $\SI{1.815}{\electronvolt}$. Spectra are background-substracted, scaled as indicated next to the trace and vertically offset for clarity. The dominant $\AEphonon$ peak is highlighted in orange. 
    (c) Logarithmic false color map of the Raman intensity as a function of excitation energy and Raman shift. Resonance with $\mathrm{X_{KK}}$ manifests as an enhanced PL background near $\SI{1.735}{\electronvolt}$ (blue arrow). Raman peaks with constant energy shift are marked by vertical arrows. Excitation energy of Raman spectra presented in (b) are marked by black horizontal arrows. 
    (d) Resonance profile of the $\AEphonon$ peak intensity for sample A (top) and sample B (bottom). Grey data points are shown with a guide to the eye. The black curve shows the calculated first-order scattering probability $P_{\mathrm{1^{st}}}$. The blue vertical line marks the energy of the exciton $E_\mathrm{X_{KK}}$ and the orange double-headed arrow indicates the energy of the $\AEphonon$ phonon.
    [Measurement conditions: Sample A: $T<\SI{300}{\milli\kelvin}$, $\SI{277}{\micro\watt}$, polarization undetermined. Sample B: $T=\SI{4}{\kelvin}$, $\SI{27}{\micro\watt}$, linear co-polarization]}
\end{figure*}

The exciton $\mathrm{X_{KK}}$ consists of a Coulomb-bound electron–hole pair with parallel spins and both carriers located at the K-point of the first Brillouin zone, and hence the parabolic exciton dispersion is centered at zero center-of-mass momentum. In addition to $\mathrm{X_{KK}}$, the established excitonic structure of WSe\textsubscript{2} illustrated in Fig. \ref{fig:Figure_A}(a) hosts several dark exciton types at comparable energies. The spin-forbidden exciton $\mathrm{X_{KK}^D}$ carries an antiparallel electron spin and lies energetically below $\mathrm{X_{KK}}$. Excitons with hole momentum at $\mathrm{K}$ and electron momentum at $\mathrm{K'}$ experience an inverted spin order in the conduction band, placing the spin-allowed exciton $\mathrm{X_{KK'}}$ below the spin-forbidden $\mathrm{X_{KK'}^D}$. Both are momentum-dark by virtue of their finite center-of-mass momentum. Additionally, the exciton $\mathrm{X_{KQ}}$, whose electron resides in the conduction band minimum at the $\mathrm{Q}$-point along the $\mathrm{\Gamma K}$ direction, is also momentum-dark. $\mathrm{X_{KQ}}$ and $\mathrm{X_{KK'}}$ are reported to lie approximately $\SIrange{50}{100}{\milli\electronvolt}$ below $\mathrm{X_{KK}}$\cite{malic_dark_2018, deilmann_finite-momentum_2019}.

The RRS results obtained from two hBN-encapsulated WSe\textsubscript{2} monolayer samples, hereafter referred to as Sample A and Sample B, are summarized in Fig. \ref{fig:Figure_A}. Four representative spectra from Sample A, spanning the measured excitation energy range, are presented in Fig. \ref{fig:Figure_A}(b). After background subtraction and suitable scaling as indicated next to each trace, several phonon peaks are resolved. The dominating peak at $\SI{251.1}{\per\centi\meter}$ $(\SI{31.1}{\milli\electronvolt})$ is assigned to a first-order Raman scattering process involving the Raman-active degenerate $\AEphonon$ optical phonon mode (orange overlay). A weaker asymmetric peak near $\SI{262}{\per\centi\meter}$ labeled 3 is commonly attributed to multiple phonon scattering including the second-order 2LA(M) Raman scattering process involving two LA(M) phonons\cite{zhao_lattice_2013, terrones_new_2014}.

At an excitation energy of $\SI{1.815}{\electronvolt}$ that is just outside the resonance with the exciton $\mathrm{X_{KK}}$, a broader peak (*) at $\SI{248}{\per\centi\meter}$ occurs in the spectrum. As the excitation energy approaches the resonance, this feature (*) vanishes and for $\SI{1.751}{\electronvolt}$ additional peaks emerge at $\SI{235}{\per\centi\meter}$ (1), $\SI{245}{\per\centi\meter}$ (2) and $\SI{268}{\per\centi\meter}$ (4). A detailed assignment of all observed peaks to specific phonon modes is provided in the Supplemental Information.

The complete RRS dataset for Sample A is shown as a logarithmic false color map in Fig. \ref{fig:Figure_A}(c), displaying the photon counts (Raman intensity) as a function of the excitation energy and the Raman shift. In addition to the phonon peaks, which appear at constant Raman shift, a broad feature centered near $\SI{1.735}{\electronvolt}$ is observed and attributed to resonantly excited PL emission from the exciton $\mathrm{X_{KK}}$ (blue arrow). The PL background has been subtracted in Fig. \ref{fig:Figure_A}(b). The map reveals a strong enhancement of the RRS intensity as the excitation energy approaches $\mathrm{X_{KK}}$ .

To quantify the resonance behavior, all spectra were fitted with a multi-Lorentzian model supplemented by a Gaussian component to account for the PL background. We now focus on the excitation-energy dependence of the $\AEphonon$ Raman peak intensity. The peak intensity extracted from the lineshape analysis yields the resonance profile shown in Fig. \ref{fig:Figure_A}(d) (upper panel), which exhibits a two peak structure. The two peaks differ in intensitiy and lineshape, with an energy separation of $\SI[separate-uncertainty=true]{33.3\pm 0.4}{\milli\electronvolt}$.

To interpret this behavior, we compare the experimental resonance profile with a theoretical model based on a first-order Raman scattering process. The Stokes process involves three steps: exciton creation by absorption of an incoming photon, $\AEphonon$ phonon emission, and exciton recombination by emission of a scattered photon. The corresponding scattering channel is illustrated by the Feynman diagram in Fig. \ref{fig:Figure_B}(a). Within third-order perturbation theory, the excitation-energy-dependent scattering probability of this process is given by Eq. \ref{eq:first_order_Raman}\cite{yu_fundamentals_2010}. 

\begin{widetext}
\begin{equation}
\label{eq:first_order_Raman}
\begin{gathered}
    P_{\mathrm{1^{st}}}(E_{\mathrm{i}})=\left|a_{\mathrm{1^{st}}}(E_{\mathrm{i}})\right|^2 = \left|\frac{\left<\mathrm{cgs}\middle|H_{\mathrm{ex.phot.}}\middle|\mathrm{X_{KK}}\right>\left<\mathrm{X_{KK}}\middle|H_{\mathrm{ex.phon.}}\middle|\mathrm{X_{KK}}\right>\left<\mathrm{X_{KK}}\middle|H_{\mathrm{ex.phot.}}\middle|\mathrm{cgs}\right>}{\left(E_{\mathrm{i}}-E_{\scriptstyle{\mathrm{A}_{1}}^{\prime}/\mathrm{E}^{\prime}}-E_{\mathrm{X_{KK}}}-\text{i}\frac{\mathit{\Gamma}_{\mathrm{X_{KK}}}}{2}\right)\left(E_{\mathrm{i}}-E_{\mathrm{X_{KK}}}-\text{i}\frac{\mathit{\Gamma}_{\mathrm{X_{KK}}}}{2}\right)}\right|^2\cdot \mathcal{F}\left(E_{\mathrm{i}}-E_{\scriptstyle{\mathrm{A}_{1}}^{\prime}/\mathrm{E}^{\prime}},E_{\mathrm{i}}\right)
\end{gathered}
\end{equation}
\end{widetext}

In this expression, a bright exciton state $\left|\mathrm{X_{KK}}\right>$ serve as the sole intermediate state, as the optical accessible exciton states are restricted by the photon momentum close to the $\Gamma$ point and is therefore approximated by a single representative state. While the exciton-phonon ($H_\mathrm{ex.phon.}$) interaction is energy independent, the energy dependence of the exciton-photon ($H_\mathrm{ex.phot.}$) interaction matrix elements is separated into a correction function $\mathcal{F}$ allowing the numerator to be treated as approximately constant. Finite exciton lifetimes are incorporated through a damping parameter $\mathit{\Gamma}_\mathrm{X_{KK}}$. Further details of the scattering probability formula are provided in the Supplemental Information. 

Within the scattering probability $P_{\mathrm{1^{st}}}$, the resonance condition arises from the denominator which compares the excitation energy $E_{\mathrm{i}}$ with the exciton energy $E_\mathrm{X_{KK}}$ before and after phonon emission, yielding an incoming and an outgoing resonance, whose energetic separation is fixed by the $\AEphonon$ phonon energy $E_{\scriptstyle{\mathrm{A}_{1}}^{\prime}/\mathrm{E}^{\prime}}=\SI{31.1}{\milli\electronvolt}$ as determined from the Raman shift. These resonances are visualized in the energy level diagrams in Fig. \ref{fig:Figure_B}(a).

The first-order model fails to reproduce the experimental observations. In particular, it enforces a peak separation equal to the phonon energy $E_{\scriptstyle{\mathrm{A}_{1}}^{\prime}/\mathrm{E}^{\prime}}$, which is smaller than the experimentally observed value. The calculated scattering probability $P_{\mathrm{1^{st}}}$ (black solid line) and the phonon energy (orange double-headed arrow) are shown in Fig. \ref{fig:Figure_A}(d). In contrast to the measured asymmetric resonance profile, the calculated two peak structure exhibits only a weak asymmetry due to the mentioned energy dependent correction function $\mathcal{F}$.

For comparison, the resonance profile of Sample B is shown in Fig. \ref{fig:Figure_A}(d) (lower panel). It also exhibits two peaks with a reduced asymmetry in the peak heights compared to Sample A, while the lineshape asymmetry persists. The peak energy separation of $\SI[separate-uncertainty=true]{35.5\pm 0.5}{\milli\electronvolt}$ exceeds that of Sample A, further deviating from the prediction of the first-order Raman model. 

\section{\label{sec:intervalley_scattering_model}Higher order intervalley scattering model and discussion}

\begin{figure*}
    \includegraphics[]{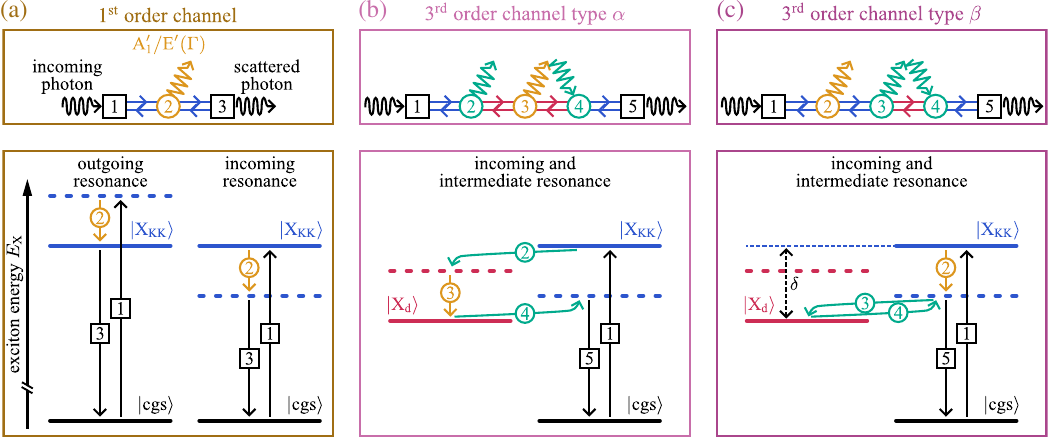}
    \caption{\label{fig:Figure_B}
    Raman scattering channels via bright and momentum dark excitons. Feynman diagrams show the order of the exciton-phonon scattering steps (circular nodes); the first and last steps represent optical excitation and radiative recombination via exciton-photon interaction (square nodes). Intermediate states are momentum-direct (blue) and momentum-indirect (red) excitons connected by bidirectional arrow. The corresponding energy level diagrams below indicate possible resonances. Resonantly populated states are shown as solid lines; non-resonant states as dashed lines. The states are the crystal ground state $\left|\mathrm{cgs}\right>$ (black), the momentum-direct exciton state $\left|\mathrm{X_{KK}}\right>$ (blue) and the momentum-indirect exciton state $\left|\mathrm{X_{d}}\right>$ (red) with the energetic detuning $\delta$. 
    (a) 1\textsuperscript{st} order channel: an $\AEphonon$ (orange) phonon is emitted from $\left|\mathrm{X_{KK}}\right>$ giving rise to an incoming and an outgoing resonance.
    (b) 3\textsuperscript{rd} order channel type $\alpha$: intervalley scattering between $\left|\mathrm{X_{KK}}\right>$ and $\left|\mathrm{X_{d}}\right>$ by emission of finite momentum phonon $\kappa$ (green). An incoming resonance with $\left|\mathrm{X_{KK}}\right>$ is followed by emission of $\kappa$ and $\AEphonon$ phonons, establishing a resonance with $\left|\mathrm{X_{d}}\right>$. Finally, a $\kappa$ phonon is absorbed.
    (c) 3\textsuperscript{rd} order channel type $\beta$: a permutation of the channel type $\alpha$ in which the $\AEphonon$ phonon is emitted from $\left|\mathrm{X_{KK}}\right>$ rather than from $\left|\mathrm{X_{d}}\right>$.
    In the Supplementary Information the 3\textsuperscript{rd} order channel type $\gamma$ is sketched, where in comparison to type $\beta$ the $\AEphonon$ is emitted after the two finite momentum scattering events.}
\end{figure*}

The discrepancy between the experimental resonance profiles and the prediction by the scattering probability $P_{\mathrm{1^{st}}}$ motivates an extension of the model by incorporating additional scattering channels mediated by momentum-dark excitons. Such excitons can be accessed from $\mathrm{X_{KK}}$ via intervalley scattering involving finite-momentum phonons, forming a higher-order Raman scattering process. 

The applied third-order scattering process is illustrated by the Feynman diagram in Fig. \ref{fig:Figure_B}(b), that includes three phonon scattering events. Two involve scattering with a finite-momentum phonon (green wavy arrows): one Stokes and one anti-Stokes scattering event that enable transitions to and from a momentum-dark exciton while conserving overall momentum. The third event is a Stokes scattering with the $\AEphonon$ phonon (orange wavy arrow), which determines the net Raman shift. Analogous third-order processes were theoretically described by Ralston \textit{et al.}\cite{ralston_resonant_1970} and subsequently applied to experimental RRS on carbon nanotubes by Haroz \textit{et al.}\cite{haroz_asymmetric_2015}. 

While the first and final intermediate states are constrained by the photon momentum, the remaining intermediate states and the involved phonons can, in principle, have arbitrary momentum. However, as established for higher-order Raman scattering in graphene, resonant enhancement results into a dominating contribution of scattering channels involving excitions and phonons with matching energy and momentum. Since the phonon energies are small compared with the exciton energies only states from the parabolic momentum-dark exciton dispersion sketched in Fig. \ref{fig:Figure_A}(a) are accessible. We consequently approximate the process by a minimal model consisting of the bright exciton state $\left|\mathrm{X_{KK}}\right>$, a single momentum-dark exciton state $\left|\mathrm{X_{d}}\right>$ and one finite-momentum phonon $\kappa$ connecting the two states. This approximation is justified because the resonantly populated states within the parabolic momentum-dark exciton dispersion have similar energies, because the phonon dispersion is nearly flat near high-symmetry points at the zone boundary, such as the $M$ and $K'$ points allowing a single representative phonon energy to capture the dominant contribution.

The three phonon scattering events of the third order scattering process can occur in six different permutations, each defining a distinct scattering sequence. Applying the assumptions described above, the intermediate excitonic states after each scattering event are uniquely determined by momentum conservation to be either $\mathrm{X_{KK}}$ or $\mathrm{X_{d}}$, such that each permutation corresponds to a unique scattering channel, resulting in six distinct third-order scattering channels.
At low temperatures, thermal population of the finite-momentum phonon is negligible. Accordingly, the three channels in which an anti-Stokes event precedes the corresponding Stokes event are neglected, leaving three contributing third-order channels.

The two most relevant channels type $\alpha$ and type $\beta$ are presented in Fig. \ref{fig:Figure_B}(b) and (c) and the third channel type $\gamma$ in the Supplemental Information. In type $\alpha$, the $\AEphonon$ phonon emission occurs between the two finite-momentum phonon scattering events with the momentum-dark exciton $\mathrm{X_{d}}$. In type $\beta$, the $\AEphonon$ phonon emission is interchanged with the finite-momentum phonon Stokes process. The third-order channel type $\gamma$ has both finite momentum phonon scattering events preceding the $\AEphonon$ phonon emission. Consequently, in type $\beta$ and type $\gamma$, the $\AEphonon$ phonon is emitted from $\mathrm{X_{KK}}$ rather than $\mathrm{X_{d}}$, resulting in modified exciton-phonon coupling strengths and altered resonance conditions. The resonance conditions of the channels are illustrated for the incoming resonance with $\mathrm{X_{KK}}$ in the energy level diagrams in Fig. \ref{fig:Figure_B}. An additional resonance involving $\mathrm{X_{d}}$ becomes accessible when the energy detuning $\delta$ between the bright and momentum-dark exciton matches the phonon energies. The resonances in type $\alpha$ and type $\beta$ also differ because the intermediate state following the first phonon scattering event is different. 

Compared to the 1\textsuperscript{st} order channel, the 3\textsuperscript{rd} order channel type $\beta$ acquires an additional phase factor arising from the two finite-momentum phonon scattering events. Since all scattering channels occur coherently in the absence of dephasing, this phase difference affects the coherent summation of the scattering amplitudes $a_{\mathrm{1^{st},3^{rd}}}$ from each channel, as expressed in Eq. \ref{eq:third_order_Raman}, yielding the total scattering probability $P_{\mathrm{1^{st}+3^{rd}}}$ for the intervalley scattering model. The resulting interference modifies the resonance lineshapes and can shift the apparent peak positions relative to the first-order model scattering probability $P_{\mathrm{1^{st}}}$. The 1\textsuperscript{st} order channel enters Eq. \ref{eq:third_order_Raman} through its amplitude $a_{\mathrm{1^{st}}}$.

\begin{equation}
    \label{eq:third_order_Raman}
    \begin{split}
    P_{\mathrm{1^{st}+3^{rd}}}\left(E_{\mathrm{i}}\right)=\left|a_{\mathrm{1^{st}}}\left(E_{\mathrm{i}}\right)+\sum_{\textrm{permut.}} a_{\mathrm{3^{rd}}}\left(E_{\mathrm{i}}\right)\right|^2 \\
    \end{split}
\end{equation}

Like $P_{\mathrm{1^{st}}}$, this expression contains the energy $E$ and damping parameter $\mathit{\Gamma}$ of the bright and momentum-dark excitons as free parameters. The phonon energies are fixed to values extracted from the Raman spectra. For the finite-momentum phonon $\kappa$, the energy was fixed to $\SI{16.4}{\milli\electronvolt}$, corresponding to a peak that appears prominently in Raman scattering spectra presented in the Supplemental Information and is tentatively assigned to a acoustic phonon mode. The complex-valued amplitude scaling parameters of each scattering channel, which encode the interaction matrix elements, are presented in the Supplemental Information.

\begin{figure}
    \includegraphics[]{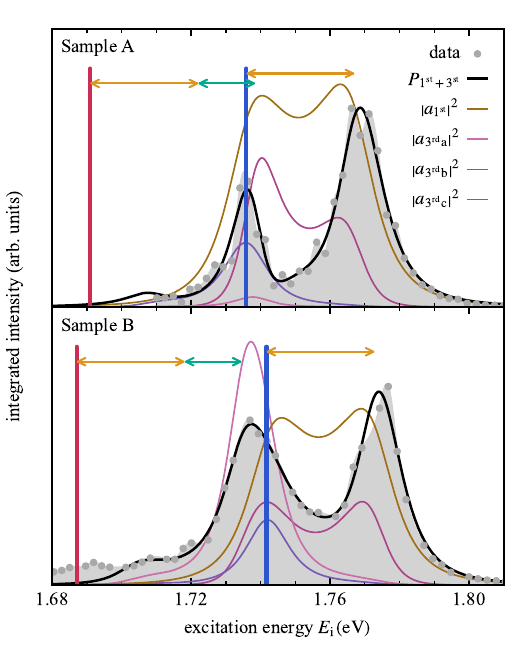}
    \caption{\label{fig:Figure_C} Description of experimental resonance profiles by the intervalley scattering model including third-order Raman scattering channels. The same data sets of Sample A (top) and Sample B (bottom) as plotted in Fig. \ref{fig:Figure_A} (d) are shown with the fitted scattering probability $P_{\mathrm{1^{st}+3^{rd}}}$ as a black line. The contributions of the individual scattering channels are shown by their squared scattering amplitudes as colored lines (cf. Fig. \ref{fig:Figure_B}).
    Blue and red vertical lines mark the energies of the bright exciton $\mathrm{X_{KK}}$ and momentum-dark exciton $\mathrm{X_{d}}$, respectively, and orange and green double headed arrows indicates the $\AEphonon$ and finite momentum phonon $\kappa$ energies, respectively, highlighting the corresponding resonance conditions. }
\end{figure}

The resonance profiles of both samples were fitted with the scattering probability $P_{\mathrm{1^{st}+3^{rd}}}$. The experimental data (grey points) and corresponding fit results (black solid lines) are shown in Fig. \ref{fig:Figure_C}), with the contributions of individual scattering channels plotted as the squared absolute values of the scattering amplitudes (colored lines).

The extracted energy parameters are summarized in Tab. \ref{tab:fitparameter}. The exciton energies $E_\mathrm{X}$, indicated by vertical lines in Fig. \ref{fig:Figure_C}), differ slightly between the samples consistent with sample-dependent effects such as strain or local dielectric variations, and remain within the range of values reported in the literature \cite{schmidt_reversible_2016, raja_coulomb_2017}. This leads to detuning $\delta$ values of $\SI{45.0\pm0.7}{\milli\electronvolt}$ and $\SI{54.7\pm1.2}{\milli\electronvolt}$ for Sample A and Sample B, respectively. The damping parameters $\mathit{\Gamma}_\mathrm{X}$ are comparable for both samples. 

\begin{table}
\caption{\label{tab:fitparameter}Energies and damping parameters obtained from fitting the scattering probability $P_{\mathrm{1^{st}+3^{rd}}}$ to the resonance profiles. *parameters determined from Raman spectra and fixed in the fitting procedure.}
\begin{ruledtabular}
\begin{tabular}{ccc}
Parameter&Sample A& Sample B\\
\hline
$E_\mathrm{X_{KK}} (\SI{}{\milli\electronvolt})$ & $1735.9\pm0.7$ & $1741.7\pm0.5$ \\
$\mathit{\Gamma}_\mathrm{X_{KK}} (\SI{}{\milli\electronvolt})$ & $21.4\pm1.5$ & $21.1\pm1.4$ \\
$E_\mathrm{X_{d}} (\SI{}{\milli\electronvolt})$ & $1690.9\pm1.0$ & $1687.0\pm1.3$ \\
$\mathit{\Gamma}_\mathrm{X_{d}} (\SI{}{\milli\electronvolt})$ & $14.9\pm1.8$ & $17.1\pm2.0$ \\
$E_{\scriptstyle{\mathrm{A}_{1}}^{\prime}/\mathrm{E}^{\prime}} (\SI{}{\milli\electronvolt})$* & $31.1$& $31.1$\\
$E_\mathrm{\kappa} (\SI{}{\milli\electronvolt})$* & $16.4$ & $16.4$ \\
\end{tabular}
\end{ruledtabular}
\end{table}

Overall, $P_{\mathrm{1^{st}+3^{rd}}}$ reproduces the lineshape of both resonance peaks and the intervening minimum well. For both samples, the 1\textsuperscript{st} order channel contributes strongly to the total scattering probability. Third-order channels also contribute significantly, although their relative weights differ between the samples. For Sample A, channels type $\beta$ and type $\gamma$ contribute moderately, while type $\alpha$ is negligible. For Sample B, type $\alpha$ contributes strongly, while type $\beta$ and type $\gamma$ contribute less.

The resonance peaks of the resonance profiles can be assigned to specific resonances within different scattering channels based on the fitted exciton energies. 
The higher-energy resonance peak is primarily associated with the outgoing resonance with the bright exciton $\mathrm{X_{KK}}$, leading to strong contribution of the  1\textsuperscript{st} order channel and the 3\textsuperscript{rd} order channel type $\beta$, which exhibits a double resonance. In the absence of a resonance with the dark exciton $\mathrm{X_{d}}$, the remaining third-order channels have only a single resonance and remain weak. In contrast, the lower-energy resonance peak arises from combination of the incoming resonance with $\mathrm{X_{KK}}$, which affects all channels, especially the third-order channel type $\gamma$, and an intermediate resonance with $\mathrm{X_{d}}$, in the third-order channel type $\alpha$ and type $\beta$. In this energy regime, all scattering channels contribute significantly and interfere, and the different detunings of the two samples give rise to markedly different lineshapes. In addition, the model predicts a weak resonance peak at approximately $\SI{1.71}{\electronvolt}$. This peak is faintly visible in the experimental data of Sample B.

As is common in higher-order Raman models with multiple coherently interfering channels, the fit of $P_{\mathrm{1^{st}+3^{rd}}}$ is not uniquely constrained by the resonance profiles alone. This ambiguity arises primarily from the large number of free parameters — in particular, the complex-valued amplitude scaling parameters encoding the interaction matrix elements. As a consequence, the relative dominance of individual third-order channels is sensitive to the chosen parameter set. Nevertheless, the extracted exciton energies $E_\mathrm{X_{KK}}$ and damping parameters $\Gamma_\mathrm{X_{KK}}$ are robustly reproduced across different parameter sets, establishing the energetic structure of the model as its physically meaningful outcome.

As introduced in Fig. \ref{fig:Figure_A}(a), several momentum-dark excitons can in principle serve as $\mathrm{X_{d}}$. To satisfy momentum conservation, the finite-momentum phonon $\kappa$ must compensate the momentum difference between $\mathrm{X_{d}}$ and $\mathrm{X_{KK}}$. Possible exciton-phonon combinations include $\mathrm{X_{KQ}}$ with a phonon near $\mathrm{-Q}$, $\mathrm{X_{KK'}}$ with a phonon near $\mathrm{K}$, and $\mathrm{X_{KQ'}}$ with a phonon near $\mathrm{M}$. The last of these excitons is not shown in Fig. \ref{fig:Figure_A}(a) as its energy lies approximately $\SI{100}{\milli\electronvolt}$ above $\mathrm{X_{KK}}$, inconsistent with the fitted detuning values \cite{deilmann_dark_2017}.

In contrast, excitons $\mathrm{X_{KQ}}$ and $\mathrm{X_{KK'}}$ have both reported detunings of approximately $\SI{50}{\milli\electronvolt}$, energetically in the vicinity of experimentally extracted detunings of $\SI{44.8}{\milli\electronvolt}$ and $\SI{54.5}{\milli\electronvolt}$\cite{deilmann_dark_2017}.
An argument favoring the assignment to $\mathrm{X_{KQ}}$ is the pronounced sample-to-sample variation of the fitted detuning energies. While the energy of $\mathrm{X_{KK'}}$ is expected to shift identical to the energy of $\mathrm{X_{KK}}$\cite{malic_dark_2018, kumar_strain_2025}, leaving the detuning largely unchanged, the energy of $\mathrm{X_{KQ}}$ is more sensitive to strain and dielectric environment \cite{schmidt_reversible_2016, raja_coulomb_2017}, consistent with the observed sample dependence. Furthermore, a scattering channel involving a spin-dark exciton appears unlikely, as phonon scattering processes in TMDCs are generally not expected to involve a spin flip \cite{gilardoni_Symmetry_2021}.

This assignment is further supported by time-resolved ARPES measurements on WSe\textsubscript{2}, which have directly observed intervalley scattering between the bright exciton $\mathrm{X_{KK}}$ and momentum-dark exciton $\mathrm{X_{KQ}}$ \cite{bange_ultrafast_2023}. While that data do not allow a distinction between electron scattering from $\mathrm{K}$ to $\mathrm{Q}$ and $\mathrm{K}$ to $\mathrm{Q'}$, it confirms the presence of efficient bright-to-dark exciton coupling on ultrafast timescales, consistent with the strong third-order channel contributions inferred from our model.

A potential limitation of the $\mathrm{X_{KQ}}$ assignment is that no pronounced phonon density of states is expected at $\mathrm{-Q}=\mathrm{-\frac{1}{2}K}$ as this momentum does not correspond to a high symmetry point of the first Brillouin zone. Consistent with this, phonons at $\mathrm{-Q}$ have, to our knowledge, not been directly reported in Raman spectroscopy experiments. In contrast, phonons at the $\mathrm{M}$ and $\mathrm{K}$ points, including the frequently discussed LA(M) mode, exhibit large phonon densities of states and are frequently observed in Raman spectra. 
 
Nevertheless, several observations support the proposed assignment. The fixed finite-momentum phonon energy $E_\kappa$ of $\SI{16.4}{\milli\electronvolt}$ is consistent with the longitudinal acoustic (LA) branch. The corresponding $LA(-Q)$ phonon exhibits in several theoretical calculated phonon dispersion a relatively flat LA branch in the vicinity of $Q$ \cite{sahin_anomalous_2013, rosner_phase_2014, jung_element-specific_2024}, which would locally enhance the phonon density of states.

The efficiency of exciton-phonon scattering, however, is not determined solely by the phonon density of states. The availability of resonant exciton states provides and additional enhancement. This is supported by theoretical calculations of excitonic linewidths that include intervalley scattering processes and predict exciton-phonon coupling involving Q-point phonons \cite{selig_excitonic_2016, brem_intrinsic_2019}. These considerations suggest that a combined exciton phonon density of states, rather then the solely phonon density of states, determines which finite-momentum phonons efficiently participate in the Raman scattering process. This interpretation remains robust regardless of the precise assignment.

At first glance, the additional scattering events in the third-order channels would be expected to substantially surpress their contribution to the total scattering probability, as also reflected in the amplitude scaling factors of the third-order channels being seven orders of magnitude smaller than those of the 1\textsuperscript{st} order channel. However, the fitted spectra indicate the opposite: the additional resonance with the momentum-dark exciton $\mathrm{X_{d}}$ effectively compensates the suppression arising from the interaction matrix elements.

The pronounced contribution of the third-order channels, in particular the channels $\beta$ and $\gamma$ demonstrates efficient coupling between the bright exciton $\mathrm{X_{KK}}$ and the momentum-dark exciton $\mathrm{X_{d}}$ by scattering with the finite-momentum phonon $\kappa$. Such strong intervalley exciton-phonon coupling suggests that the bright and momentum-dark excitons cannot be regarded as independent excitations but instead interact coherently whenever their energy detuning is sufficiently small, significant dephasing is absent and a suitable phonon mode connects the excitons.

In this picture, the momentum-dark exciton $\mathrm{X_{d}}$ together with the finite momentum phonon $\kappa$ forms a composite excitation $\mathrm{X_{d}}+\kappa$. Because the phonon compensates the exciton momentum, the total momentum of the composite excitation is approximately zero, allowing direct optical coupling and therefore effectively brightening the momentum-dark exciton. Similar concepts have been discussed as phonon assisted emission of momentum-dark excitons\cite{brem_phonon-assisted_2020} and phonon-dressed dark exciton states\cite{chand_phonondressed_2026}. Here, we extend this picture to a coupled phonon-dressed bright-dark exciton state.

When the energy of this composite excitation $\mathrm{X_{d}}+\kappa$ approaches - for example strain or doping induced - that of the bright exciton $\mathrm{X_{KK}}$ the coupling can hybridize the two excitations. The resulting eigenstates are linear combinations of both excitations, leading to modified energies and oscillator strengths. These states could be applied in a first-order Raman scattering process. In literature, emission features in strain dependent PL have previously been interpreted in terms of hybridized excitons \cite{kumar_strain_2024,kumar_strain_2025}.

It should be emphasized, however, that both $\mathrm{X_{KK}}$ and $\mathrm{X_{d}}$ represent dispersive exciton continua rather than isolated states, as assumed in the intervalley scattering model discussed above. Consequently, the linear combination is not restricted to two discrete excitations but involves an extended manifold of exciton states coupled to multiple phonon branches throughout momentum space. The two level linear combination discussed here should therefore be regarded as an intuitive minimal model. A microscopic theory accounting for the full exciton and phonon dispersions could quantify the resulting exciton energy shifts and the modified optical response, but we leave a quantitative treatment of these effects to future studies.

Although the intervalley scattering model discussed in detail and the linear combination picture provide different theoretical descriptions, they represent complementary views of the same underlying exciton-phonon interaction. This underscores the important role of the momentum-dark excitons in explaining the optical properties of WSe\textsubscript{2}. More broadly, the proposed mechanism provides a a possible explanation for peculiar optical properties reported for WSe\textsubscript{2} monolayers: phonon-brightened momentum-dark excitons provide a natural explanation for emission features observed below the bright exciton energy, while phonon-dressed linear combinations of bright and momentum-dark exciton states may contribute to the reduced valley polarization. Extending this concept to bilayer and heterostructure TMDCs further suggests that analogous phonon-assisted linear combinations of excitons — or equivalently, coupled phonon-dressed bright-dark exciton states — could provide efficient pathways for interlayer charge transfer.
 
\section{\label{sec:conclusion}Conclusion}

In conclusion, we investigated the resonant Raman scattering on hBN encapsulated WSe\textsubscript{2} monolayers at cryogenic temperatures by continuously tuning the excitation energy across the bright exciton $\mathrm{X_{KK}}$ resonance. The resonance profiles of the $\AEphonon$ phonon mode exhibit two distinct resonance peaks whose energy separation significantly exceeds the $\AEphonon$ phonon energy. While a first-order Raman scattering model predicts two resonance peaks separated by the phonon energy, it fails to reproduce the observed asymmetric lineshape, unequal peak intensities, and the anomalous peak separation. These discrepancies are resolved by extending the model to include coherent third-order scattering channels, in which finite-momentum phonons facilitate intervalley coupling between the bright exciton $\mathrm{X_{KK}}$ and a momentum-dark exciton $\mathrm{X_{d}}$. Based on the extracted detuning energies and their sample-to-sample variation, we tentatively assign the relevant dark exciton to the $\mathrm{X_{KQ}}$ exciton and the finite-momentum phonon to the $LA$ phonon branch near $\mathrm{-Q}$, consistent with the possibility of double and triple resonance enhancement of the higher-order scattering channels. We additionally propose a description in terms of phonon-dressed linear combinations bright and momentum-dark exciton or equivalently phonon-hybridized bright-dark exciton states participating in effective first-order Raman scattering processes. Taken together, our results establish resonance profile analysis as a sensitive probe of momentum-dark exciton energetics and exciton-phonon coupling in WSe\textsubscript{2}, and provide a microscopic perspective on the material's anomalous optical properties.

\appendix

\section{Acknowledgements}

The authors gratefully acknowledge financial support by the German Science Foundation (DFG) via Grants No.443274199 and 556436549 (WU 637/7-2,8-1), HO3324/13-1, HO3324/16-1 and the priority program 2244 (2DMP) as well as MCQST, and e-conversion.

We thank Daniel Groll, Daniel Wigger and Tilmann Kuhn for extensive discussions.

\section{Data availability statement}

The data that support the findings of this study are available from the corresponding author upon reasonable request.

\section{Declaration of generative AI and AI-assisted technologies}

During the preparation of this work, the authors used large language models to refine the manuscript regarding grammar, readability and clarity. After using these tools, the authors reviewed and edited the content as needed and take full responsibility for the content of the publication. No additional use of artificial intelligence was made during data acquisition, analysis, interpretation, or other aspects of this study.

%

\clearpage
\pagebreak

\widetext

\begin{center}
\textbf{\large Supplemental Information to:\\
Resonant Raman signatures of bright and momentum-dark exciton coupled by intervalley phonon scattering in monolayer WSe\textsubscript{2}}
\end{center}

\setcounter{equation}{0}
\setcounter{figure}{0}
\setcounter{table}{0}
\setcounter{page}{1}
\setcounter{section}{0}

\renewcommand{\theequation}{S\arabic{equation}}
\renewcommand{\thefigure}{S\arabic{figure}}
\renewcommand{\bibnumfmt}[1]{[S#1]}
\renewcommand{\citenumfont}[1]{S#1}
\renewcommand{\thepage}{S\arabic{page}}
\renewcommand{\thesection}{S\Roman{section}}

\section{Details on sample fabrication}

Both samples consisted of a monolayer of WSe\textsubscript{2} encapsulated between a top and bottom hexagonal boron nitride (hBN) flakes and placed on a Si/SiO\textsubscript{2} substrate. Sample A was prepared using the PDMS method, whereas Sample B was prepared using the PC method. In both cases, monolayer WSe\textsubscript{2} flakes were mechanically exfoliated from commercially available bulk WSe\textsubscript{2} crystals (HQ Graphene) using adhesive tape. For Sample A, the exfoliated flakes were transferred onto a PDMS stamp supported by a glass slide, whereas for Sample B they were transferred onto a Si/SiO\textsubscript{2} substrate. Monolayer flakes were identified by optical microscopy based on their optical contrast. Subsequently, both samples were assembled into fully hBN encapsulated heterostructures using a home-built stacking setup. For Sample A, the individual flakes were sequentially stamped onto the substrate with wet chemical cleaning and annealing in an Ar/N\textsubscript{2} atmosphere at \SI{200}{\degreeCelsius} and \SI{20}{\milli\bar} for \SI{30}{\minute} between each stacking step. For Sample B, the top hBN and selected WSe\textsubscript{2} monolayer were picked up from the Si/SiO\textsubscript{2} substrate using a PC stamp and finally released onto the bottom hBN flake prepared on the Si/SiO\textsubscript{2} substrate. Residual PC on the stack was removed by immersing the sample in chloroform for several hours. Afterwards, the sample was annealed in vacuum (\SI{1e-4}{\milli\bar}) at \SI{200}{\degreeCelsius} for \SI{12}{\hour}.

\section{Detailed formulas of the 1st order Raman and intervalley scattering model}
\label{sec:detailed formulas}

The microscopic-Raman-scattering model employed in this work is derived following the microscopic Raman theory presented in the textbook \textit{Fundamentals of Semiconductors: Physics and Materials Properties}  by P. Y. Yu and M. Cardona \cite[Chap.~7.2.5]{yu_fundamentals_2010}. 

The Raman scattering probability for a phonon ph in a first-order Raman scattering process is derived using third order perturbation theory. The three perturbations correspond to a exciton creation by photon absorption via exciton-photon interaction $H'_\mathrm{ex.phot.}$, phonon creation by exciton scattering via exciton-phonon interaction $H_\mathrm{ex.phon.}$ and exciton annihilation by emission of the scattered photon via exciton-photon interaction $H'_\mathrm{ex.phot.}$. In general, the scattering events could occur in arbitrary scattering sequences. However, scattering sequences that do not start with photon absorption and do not end with emission of the scattered photon are highly off-resonant because of the large difference between the exciton and phonon energy. Therefore, theses off-resonant sequences are neglected in the following. For the first-order Raman scattering process the resulting Eq. \ref{eq:first_order_Raman_general} contains a single scattering sequence in the same order as the scattering events introduced above.

\begin{equation}
\label{eq:first_order_Raman_general}
    P(E_\mathrm{i}) =\left|\sum_\mathrm{n,m}\frac{\left<\mathrm{cgs}|H'_{\mathrm{ex.phot.}}|\mathrm{m}\right>\left<\mathrm{m}|H_{\mathrm{ex.phon.}}|\mathrm{n}\right>\left<\mathrm{n}|H'_{\mathrm{ex.phot.}}|\mathrm{cgs}\right>}{\left(E_\mathrm{i}-E_{\mathrm{ph}}-E_\mathrm{m}-\text{i}\frac{\Gamma_\mathrm{m}}{2}\right)\left(E_\mathrm{i}-E_\mathrm{n}-\text{i}\frac{\Gamma_\mathrm{n}}{2}\right)}\right|^2
\end{equation}

The nominator of this formula contains the scattering matrix elements of every included scattering event, while the denominator contains one term per intermediate state that compares the energy of the intermediate state with the excitation energy after reduction of energy by emission of phonons. The width of the intermediate state is approximated by the full width half maxima value $\Gamma$. The sums over $m$ and $n$ run over all accessible excited states. 
As the bright exciton $\mathrm{X}_\mathrm{KK}$ is the only accessible intermediate state within the considered excitation-energy range, the sum can be evaluated explicitly, resulting in a single scattering channel. When the $\AEphonon$ phonon is involved in the scattering process $\left(E_{\mathrm{ph}}=E_{AE}\right)$, the expression for the first-order-Raman scattering probability simplifies to Eq. \ref{eq:first_order_Raman_SI}. The corresponding scattering amplitude $a_{\mathrm{1^{st}}}$ is defined accordingly.

\begin{equation}
\label{eq:first_order_Raman_SI}
\begin{gathered}
    P_{\mathrm{1^{st}}}(E_{\mathrm{i}})=\left|a_{\mathrm{1^{st}}}(E_{\mathrm{i}})\right|^2 = \left|\frac{\left<\mathrm{cgs}\middle|H'_{\mathrm{ex.phot.}}\middle|\mathrm{X_{KK}}\right>\left<\mathrm{X_{KK}}\middle|H_{\mathrm{ex.phon.}}^{\mathrm{AE}}\middle|\mathrm{X_{KK}}\right>\left<\mathrm{X_{KK}}\middle|H'_{\mathrm{ex.phot.}}\middle|\mathrm{cgs}\right>}{\left(E_{\mathrm{i}}-E_{\mathrm{AE}}-E_{\mathrm{X_{KK}}}-\text{i}\frac{\Gamma_{\mathrm{X_{KK}}}}{2}\right)\left(E_{\mathrm{i}}-E_{\mathrm{X_{KK}}}-\text{i}\frac{\Gamma_{\mathrm{X_{KK}}}}{2}\right)}\right|^2
\end{gathered}
\end{equation}

The extension to the third-order Raman scattering process increases the perturbation theory to fifth order. In the general expression for the scattering probability, a summation over four intermediate states is required, and different phonon scattering events may contribute. As discussed in the main text, the model is restricted to the bright exciton $\mathrm{X}_\mathrm{KK}$ and a momentum-dark exciton $\mathrm{X}_\mathrm{d}$ as intermediate exciton states, together with three phonon-scattering events: a Stokes-scattering event involving the $\AEphonon$ phonon, a Stokes scattering event involving the finite-momentum phonon $\kappa$ and an anti-Stokes scattering event involving the finite-momentum phonon $\kappa$. To keep the complexity of the model manageable, only a single momentum-dark exciton $\mathrm{X}_\mathrm{d}$ and a single finite-momentum phonon $\kappa$ are considered. Possible assignments are discussed in the main text.

The three phonon-scattering events can occur in six different scattering sequences, while exciton and re-emission are still fixed to the first and last scattering steps, respectively. Each scattering sequence corresponds to one scattering channel, as the intermediate states are uniquely determined by momentum conservation. As discussed in the main text, we neglect scattering channels in which the anti-Stokes scattering of $\kappa$ occurs before the corresponding Stokes-scattering event, because the low sample temperature prevents thermal population of finite-momentum phonons, regardless of whether they belong to the acoustic or optical branch. The scattering amplitudes of the three remaining third-order scattering channels are added coherently together with the first-order Raman scattering channel amplitude, thereby allowing for interference effects. The excitation-energy-dependent scattering amplitudes are given in Eq. \ref{eq:detailed_third-order}, where the complex amplitude scaling parameters $a$, $b$ and $c$ are introduced as defined in Eq. \ref{eq:amplitude_scaling_parameter}. These parameters are treated as fitting parameters during the fitting procedure.

\begin{equation}
     P_{\mathrm{1^{st}+3^{rd}}}=\left|a_{\mathrm{1^{st}}}+a_{\mathrm{3^{rd}},\alpha}+a_{\mathrm{3^{rd}},\beta}+a_{\mathrm{3^{rd}},\gamma}\right|^2
\end{equation}
    
{\fontsize{8}{9}\begin{align}
    \label{eq:detailed_third-order}
    \begin{split}
    a_{\mathrm{1^{st}}}&=
    \frac{\left<\mathrm{cgs}\middle|H'_{\mathrm{ex.phot.}}\middle|\mathrm{X_{KK}}\right>\left<\mathrm{X_{KK}}\middle|H_{\mathrm{ex.phon.}}^{\mathrm{AE}}\middle|\mathrm{X_{KK}}\right>\left<\mathrm{X_{KK}}\middle|H'_{\mathrm{ex.phot.}}\middle|\mathrm{cgs}\right>}
    {\left(E_{\mathrm{i}}-E_{\mathrm{AE}}-E_{\mathrm{X_{KK}}}-\text{i}\frac{\Gamma_{\mathrm{X_{KK}}}}{2}\right)\left(E_{\mathrm{i}}-E_{\mathrm{X_{KK}}}-\text{i}\frac{\Gamma_{\mathrm{X_{KK}}}}{2}\right)}\\
    &=\frac{a}
    {\left(E_{\mathrm{i}}-E_{\mathrm{AE}}-E_{\mathrm{X_{KK}}}-\text{i}\frac{\Gamma_{\mathrm{X_{KK}}}}{2}\right)\left(E_{\mathrm{i}}-E_{\mathrm{X_{KK}}}-\text{i}\frac{\Gamma_{\mathrm{X_{KK}}}}{2}\right)}\\
    \\
    a_{\mathrm{3^{rd}},\alpha}&=
    \frac{\left<\mathrm{cgs}\middle|H'_{\mathrm{ex.phot.}}\middle|\mathrm{X_{KK}}\right>\left<\mathrm{X_{KK}}\middle|H_{\mathrm{ex.phon.}}^{\kappa}\middle|\mathrm{X_{d}}\right>\left<\mathrm{X_{d}}\middle|H_{\mathrm{ex.phon.}}^{\mathrm{AE}}\middle|\mathrm{X_{d}}\right>\left<\mathrm{X_{d}}\middle|H_{\mathrm{ex.phon.}}^{\mathrm{\kappa}}\middle|\mathrm{X_{KK}}\right>\left<\mathrm{X_{KK}}\middle|H'_{\mathrm{ex.phot.}}\middle|\mathrm{cgs}\right>}
    {\left(E_{\mathrm{i}}-E_{\mathrm{AE}}-E_{\mathrm{X_{KK}}}-\text{i}\frac{\Gamma_{\mathrm{X_{KK}}}}{2}\right)\left(E_{\mathrm{i}}-E_{\mathrm{\kappa}}-E_{\mathrm{AE}}-E_{\mathrm{X_{d}}}-\text{i}\frac{\Gamma_{\mathrm{X_{d}}}}{2}\right)\left(E_{\mathrm{i}}-E_{\mathrm{\kappa}}-E_{\mathrm{X_{d}}}-\text{i}\frac{\Gamma_{\mathrm{X_{d}}}}{2}\right)\left(E_{\mathrm{i}}-E_{\mathrm{X_{KK}}}-\text{i}\frac{\Gamma_{\mathrm{X_{KK}}}}{2}\right)}\\
    &=\frac{a \cdot b \cdot c}
    {\left(E_{\mathrm{i}}-E_{\mathrm{AE}}-E_{\mathrm{X_{KK}}}-\text{i}\frac{\Gamma_{\mathrm{X_{KK}}}}{2}\right)\left(E_{\mathrm{i}}-E_{\mathrm{\kappa}}-E_{\mathrm{AE}}-E_{\mathrm{X_{d}}}-\text{i}\frac{\Gamma_{\mathrm{X_{d}}}}{2}\right)\left(E_{\mathrm{i}}-E_{\mathrm{\kappa}}-E_{\mathrm{X_{d}}}-\text{i}\frac{\Gamma_{\mathrm{X_{d}}}}{2}\right)\left(E_{\mathrm{i}}-E_{\mathrm{X_{KK}}}-\text{i}\frac{\Gamma_{\mathrm{X_{KK}}}}{2}\right)}\\
    \\
    a_{\mathrm{3^{rd}},\beta}&=
    \frac{\left<\mathrm{cgs}\middle|H'_{\mathrm{ex.phot.}}\middle|\mathrm{X_{KK}}\right>\left<\mathrm{X_{KK}}\middle|H_{\mathrm{ex.phon.}}^{\mathrm{\kappa}}\middle|\mathrm{X_{KK}}\right>\left<\mathrm{X_{KK}}\middle|H_{\mathrm{ex.phon.}}^{\mathrm{\kappa}}\middle|\mathrm{X_{d}}\right>\left<\mathrm{X_{d}}\middle|H_{\mathrm{ex.phon.}}^{\mathrm{AE}}\middle|\mathrm{X_{KK}}\right>\left<\mathrm{X_{KK}}\middle|H'_{\mathrm{ex.phot.}}\middle|\mathrm{cgs}\right>}
    {\left(E_{\mathrm{i}}-E_{\mathrm{AE}}-E_{\mathrm{X_{KK}}}-\text{i}\frac{\Gamma_{\mathrm{X_{KK}}}}{2}\right)\left(E_{\mathrm{i}}-E_{\mathrm{AE}}-E_{\mathrm{\kappa}}-E_{\mathrm{X_{d}}}-\text{i}\frac{\Gamma_{\mathrm{X_{d}}}}{2}\right)\left(E_{\mathrm{i}}-E_{\mathrm{AE}}-E_{\mathrm{X_{KK}}}-\text{i}\frac{\Gamma_{\mathrm{X_{KK}}}}{2}\right)\left(E_{\mathrm{i}}-E_{\mathrm{X_{KK}}}-\text{i}\frac{\Gamma_{\mathrm{X_{KK}}}}{2}\right)}\\
    &=\frac{a \cdot c}
    {\left(E_{\mathrm{i}}-E_{\mathrm{AE}}-E_{\mathrm{X_{KK}}}-\text{i}\frac{\Gamma_{\mathrm{X_{KK}}}}{2}\right)\left(E_{\mathrm{i}}-E_{\mathrm{AE}}-E_{\mathrm{\kappa}}-E_{\mathrm{X_{d}}}-\text{i}\frac{\Gamma_{\mathrm{X_{d}}}}{2}\right)\left(E_{\mathrm{i}}-E_{\mathrm{AE}}-E_{\mathrm{X_{KK}}}-\text{i}\frac{\Gamma_{\mathrm{X_{KK}}}}{2}\right)\left(E_{\mathrm{i}}-E_{\mathrm{X_{KK}}}-\text{i}\frac{\Gamma_{\mathrm{X_{KK}}}}{2}\right)}\\
    \\
    a_{\mathrm{3^{rd}},\gamma}&=
    \frac{\left<\mathrm{cgs}\middle|H'_{\mathrm{ex.phot.}}\middle|\mathrm{X_{KK}}\right>\left<\mathrm{X_{KK}}\middle|H_{\mathrm{ex.phon.}}^{\mathrm{AE}}\middle|\mathrm{X_{d}}\right>\left<\mathrm{X_{d}}\middle|H_{\mathrm{ex.phon.}}^{\mathrm{\kappa}}\middle|\mathrm{X_{KK}}\right>\left<\mathrm{X_{KK}}\middle|H_{\mathrm{ex.phon.}}^{\mathrm{\kappa}}\middle|\mathrm{X_{KK}}\right>\left<\mathrm{X_{KK}}\middle|H'_{\mathrm{ex.phot.}}\middle|\mathrm{cgs}\right>}
    {\left(E_{\mathrm{i}}-E_{\mathrm{AE}}-E_{\mathrm{X_{KK}}}-\text{i}\frac{\Gamma_{\mathrm{X_{KK}}}}{2}\right)\left(E_{\mathrm{i}}-E_{\mathrm{X_{KK}}}-\text{i}\frac{\Gamma_{\mathrm{X_{KK}}}}{2}\right)\left(E_{\mathrm{i}}-E_{\mathrm{\kappa}}-E_{\mathrm{X_{d}}}-\text{i}\frac{\Gamma_{\mathrm{X_{d}}}}{2}\right)\left(E_{\mathrm{i}}-E_{\mathrm{X_{KK}}}-\text{i}\frac{\Gamma_{\mathrm{X_{KK}}}}{2}\right)}\\
    &=\frac{a \cdot c}
    {\left(E_{\mathrm{i}}-E_{\mathrm{AE}}-E_{\mathrm{X_{KK}}}-\text{i}\frac{\Gamma_{\mathrm{X_{KK}}}}{2}\right)\left(E_{\mathrm{i}}-E_{\mathrm{X_{KK}}}-\text{i}\frac{\Gamma_{\mathrm{X_{KK}}}}{2}\right)\left(E_{\mathrm{i}}-E_{\mathrm{\kappa}}-E_{\mathrm{X_{d}}}-\text{i}\frac{\Gamma_{\mathrm{X_{d}}}}{2}\right)\left(E_{\mathrm{i}}-E_{\mathrm{X_{KK}}}-\text{i}\frac{\Gamma_{\mathrm{X_{KK}}}}{2}\right)}
     \end{split}
\end{align}}

\begin{align}
\begin{split}
\label{eq:amplitude_scaling_parameter}
a&=\left<\mathrm{cgs}\middle|H_{\mathrm{ex.phot.}}\middle|\mathrm{X_{KK}}\right>\left<\mathrm{X_{KK}}\middle|H_{\mathrm{ex.phon.}}^{\mathrm{AE}}\middle|\mathrm{X_{KK}}\right>\left<\mathrm{X_{KK}}\middle|H_{\mathrm{ex.phot.}}\middle|\mathrm{cgs}\right>\\
b&=\frac{\left<\mathrm{X_{d}}\middle|H_{\mathrm{ex.phon.}}^{\mathrm{AE}}\middle|\mathrm{X_{d}}\right>}{\left<\mathrm{X_{KK}}\middle|H_{\mathrm{ex.phon.}}^{\mathrm{AE}}\middle|\mathrm{X_{KK}}\right>}\\
c&=\left<\mathrm{X_{KK}}\middle|H_{\mathrm{ex.phon.}}^{\mathrm{\kappa}}\middle|\mathrm{X_{d}}\right>\left<\mathrm{X_{d}}\middle|H_{\mathrm{ex.phon.}}^{\mathrm{\kappa}}\middle|\mathrm{X_{KK}}\right>\\
\end{split}
\end{align}

Finally, the excitation-energy dependence of the exciton-photon interaction $H'_\mathrm{ex.phot.}$ is factorized into an excitation-energy-independent exciton-photon interaction $H_\mathrm{ex.phot.}$ and a correction function $\mathcal{F}$ as shown in Eq. \ref{eq:correction-function}. The excitation-energy dependency originates from the energy dependence of the power emitted from a dipole, which introduces a factor $E_\mathrm{out}^4$. Since the Raman-scattering probability is formulated in terms of quantized excitations (photons and excitons) rather than the optical power of the electromagnetic field, this factor is reformulation into the correction factor $\mathcal{F}\left(E_{\mathrm{i}}-E_\mathrm{ph},E_{\mathrm{i}}\right)=E_\mathrm{out}^3\cdot E_\mathrm{in}$ \cite[Chap.~2.1.7]{cardona_light_1982}. Consequently, the scattering probability can be rewritten as the product of the scattering probability using the excitation-independent interaction $H_\mathrm{ex.phot.}$ and the correction factor, as the matrix elements involving the exciton-photon interaction are indentical for every scattering channel.

\begin{align}
\begin{split}
    \label{eq:correction-function}
    &\left|\left<\mathrm{cgs}\middle|H'_{\mathrm{ex.phot.}}\left(E_\mathrm{i}-E_\mathrm{ph}\right)\middle|\mathrm{X_{KK}}\right>\left<\mathrm{X_{KK}}\middle|H'_{\mathrm{ex.phot.}}\left(E_\mathrm{i}\right)\middle|\mathrm{cgs}\right>\right|^2\\
    &=\left|\left<\mathrm{cgs}\middle|H_{\mathrm{ex.phot.}}\middle|\mathrm{X_{KK}}\right>\left<\mathrm{X_{KK}}\middle|H_{\mathrm{ex.phot.}}\middle|\mathrm{cgs}\right>\right|^2\cdot\left(E_\mathrm{i}-E_\mathrm{AE}\right)^3 E_\mathrm{i}
\end{split}
\end{align}

\section{\label{sec:lineshapeanalysis} Details on lineshape analysis}

\begin{figure*}[h]
    \includegraphics[]{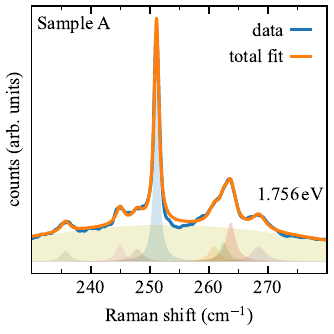}
    \caption{\label{fig:example_fit} 
   Lineshape analysis using a suitable sum of Lorentzian- and Gaussian-functions on example of an individual Raman spectrum of Sample A with an exciton energy of $\SI{1.756}{\electronvolt}$. The measured spectrum is plotted in blue and the total fit in orange. The contributions of the individual Lorentzian-functions and the broad Gaussian function are shown as filled areas.}
\end{figure*}

The measured Raman spectra are fitted individually using a multi-Lorentzian model to determine the changes in peak intensities with varying excitation energy. Some peaks exhibit slight shifts in their positions or are observable only under resonant conditions. In addition, asymmetric Raman modes require a varying number of Lorentzian functions to be adequately described, while the spectra are superimposed on a broad background. Consequently, constructing a consistent fitting model for the entire measurement series is challenging. For the results presented in the main body of the paper, only the intensity of the dominant peak assigned to the $\AEphonon$ phonon mode is required. Therefore, the fitted Raman shift range is restricted to a spectral region around the $\AEphonon$ peak, where the background contribution, tentatively assigned to photoluminescence, is approximated by a Gaussian function.

The main $\AEphonon$ peak is fitted by a single Lorentzian function. The additional peaks labeled in the main text as peaks $1$, $2$ and $4$ are each modeled by a single Lorentzian function, whereas peak $3$, which is often assigned to the 2LA(M) mode in the literature, is modeled by three Lorentzian functions. An additional Lorentzian function is included on the lower Raman shift shoulder of the $\AEphonon$ peak to describe a contribution that is not resolved as an individual peak.

For sample B, the fitting model is extended to account for polarization-resolved linearly co- and cross-polarized spectra that have been recorded. Since the $\AEphonon$ peak exhibits partial linear co-polarization, the co-polarized spectra are used to determine the resonance trace, while the linear cross-polarized spectra are included to obtain more reliable fit results. For every excitation energy, both spectra are fitted simultaneously. The cross-polarized spectrum is assigned an overall scaling factor to account for the detection efficiency differences between the two polarization configurations. In addition, individual scaling factors are applied to the amplitudes of the fitted Lorentzian functions to account for differences in the polarization dependence of the Raman modes.

\section{\label{sec:MFR} Resonant Raman spectra accessing acoustic phonon modes}

\begin{figure*}[h]
    \includegraphics[]{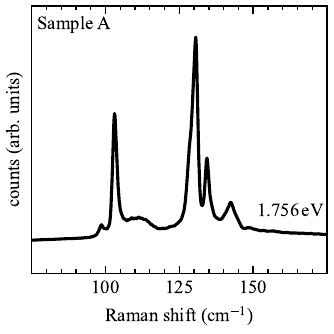}
    \caption{\label{fig:MFR}
    Raman spectra of WSe\textsubscript{2} recorded at Raman shifts for energies lower than the $\AEphonon$ phonon. The excitation energy of $\SI{1.756}{\electronvolt}$ is close to the one of the bright exciton $\mathrm{X_{KK}}$ resonantly enhancing the signal. The dominant peak in the shown spectra occurs at a Raman shift of $\SI{132.2}{\per\centi\meter}=\SI{16.4}{\milli\electronvolt}$}
\end{figure*}

The Raman scattering signal of the WSe\textsubscript{2} monolayer presented in Fig. \ref{fig:MFR} contains a group of peaks with lower energy Raman shifts. Their energies are consistent with those of the acoustic phonon branches of WSe\textsubscript{2} at finite-momentum \cite{sahin_anomalous_2013}, excluding the $\Gamma$ point where the acoustic phonon energies become vanishingly small. These Raman peaks are therefore tentatively assigned to defect-assisted Raman scattering processes involving an quasi-elastic defect-scattering event and a finite-momentum acoustic phonon scattering. Accordingly, the energy of each Raman mode is assigned to that of a corresponding finite-momentum acoustic phonon. Similar defect-assisted Raman scattering processes are well known in graphene, where they give rise to the D peak \cite{cancado_Stokes_2002, malard_raman_2009} and have also been reported for other TMDCs \cite{carvalho_intervalley_2017}. 

In the main body of the paper, we used the dominant peak at $\SI{16.4}{\milli\electronvolt} (\SI{132.2}{\per\centi\meter})$ to fix the energy of the finite-momentum phonon $\kappa$ in the third-order Raman scattering process. The observed Raman peaks exhibit only small energy differences, for example the second most intense peak corresponds to an energy of $\SI{13.0}{\milli\electronvolt} (\SI{104.9}{\per\centi\meter})$. Consequently, if a different peak within this group would be assigned to the finite-momentum acoustic phonon involved in the third-order Raman scattering process, our analysis would be affected only to a minor extend.

\section{Assignment of experimental Raman modes}

\begin{table}[h]
\caption{\label{tab:peak assignment}Assignment of Raman modes observed in the spectra of Sample A excited at an energy of \SI{1.756}{\electronvolt}. The experimental peak positions were extracted from the lineshape analysis presented in \ref{sec:lineshapeanalysis}. The peaks are tentatively assigned to phonons modes according to their energies. The theoretical phonon frequencies and the label of the phonon mode according to a phonon branch are taken from Ref. \cite{de_luca_new_2020}.}
\begin{ruledtabular}
\begin{tabular}{cccc}
Label in Fig. 1 & Experimental frequency &  Tentative assignment & Theoretical frequency \\
\hline
$a$ & $\SI{235.7}{\per\centi\meter}$ & TO\textsubscript{2}(M) & $\SI{229.1}{\per\centi\meter}$\\
\hline
$b$ & $\SI{245.0}{\per\centi\meter}$ &  LO\textsubscript{2}(M) & $\SI{245.2}{\per\centi\meter}$\\
\hline
$*$ & $\SI{247.9}{\per\centi\meter}$ & & \\
\hline
$\AEphonon$ & $\SI{251.1}{\per\centi\meter}$ & ${\mathrm{E}}^{\prime}$=LO\textsubscript{2}/TO\textsubscript{2}($\Gamma$) & $\SI{250.8}{\per\centi\meter}$\\
 &  & ${\mathrm{A}_{1}}^{\prime}$=ZO\textsubscript{2}($\Gamma$) & $\SI{251.3}{\per\centi\meter}$\\
\hline
$c$ &  & ZO\textsubscript{1}(K) & $\SI{257.3}{\per\centi\meter}$\\
 &  & LO\textsubscript{2}(K)& $\SI{258.9}{\per\centi\meter}$\\
 & $\SI{260.9}{\per\centi\meter}$ & 2LA(M) & $\SI{260.2}{\per\centi\meter}$\\
 & $\SI{262.5}{\per\centi\meter}$ & ZO\textsubscript{2}(K) & $\SI{261.9}{\per\centi\meter}$\\
 & $\SI{263.7}{\per\centi\meter}$ & ZO\textsubscript{1}(M) & $\SI{263.3}{\per\centi\meter}$\\
\hline
$d$ & $\SI{268.4}{\per\centi\meter}$ & ZA(K)+LA(K)& $\SI{266.1}{\per\centi\meter}$\\
 &  & ZO\textsubscript{2}(M) & $\SI{270.0}{\per\centi\meter}$\\
\end{tabular}
\end{ruledtabular}
\end{table}

Although the main focus of this work is the $\AEphonon$ phonon mode, several additional Raman modes appear in the spectra. Their frequencies, obtained from the lineshape analysis, together with tentative mode assignment are summarized in Tab. \ref{tab:peak assignment}. The assignment of Raman modes in monolayer WSe\textsubscript{2} to specific phonon modes has been the subject of numerous previous studies \cite{del_corro_excited_2014, terrones_new_2014, de_luca_new_2020, mcdonnell_observation_2020, blaga_unveiling_2024}. The mode assignments are based on the comparison between experimentally observed peak energies and DFT calculated phonon dispersion. Polarization-resolved Raman measurements can provide further evidence through comparisons with the Raman tensors of the candidate modes. Nevertheless, the assignments of several Raman peaks remain ambiguous in the literature and noticeable differences between reports persist. This is partly because the observed Raman spectra depend on the sample conditions like temperature\cite{dos_santos_temperature_2020}, strain\cite{sahin_anomalous_2013, dadgar_strain_2018, chaurasiya_strain-mediated_2018}, defect density \cite{mignuzzi_effect_2015, shi_raman_2016} and scattering geometry. In addition, the excitation energy allows mode-specific resonant excitation \cite{del_corro_excited_2014, mcdonnell_observation_2020}. Furthermore, theoretically derived phonon dispersion relations also show discrepancies between different reports\cite{sahin_anomalous_2013, rosner_phase_2014, jung_element-specific_2024}.

Usually three types of Raman scattering processes are considered for the assignment of the observed Raman peaks: (i) first-order Raman scattering processes involving phonons with momenta close to the center of the first Brillouin zone ($\Gamma$); (ii) second-order Raman scattering processes involving two phonon scattering events that contribute with opposite momenta, thereby fulfilling momentum conservation. This category comprises both addition modes (including overtones) and difference modes. At low sample temperatures, difference modes are usually not considered because they involve anti-Stokes scatting events, which are less probable due to the reduced phonon population; (iii) defect-assisted Raman scattering process that combine a phonon scattering event with a quasi-elastic defect scattering event, thereby relaxing the momentum conservation requirement. Both defect-assisted and second-order process allow finite-momentum phonons to contribute, providing access to additional points in the phonon dispersion with high density of state. In contrast to the second-order Raman scattering process, the third order Raman scattering process discussed in the main text is less affect by low phonon occupation because the preceding Stokes event increases the population of the finite-momentum phonon $\kappa$ before the subsequent anti-Stokes scattering event occurs.

The assignment of the most intense Raman mode at $\SI{251.1}{\per\centi\meter}$ to the degenerate ${\mathrm{A}_{1}}^{\prime}/{\mathrm{E}}^{\prime}$ mode is supported by its partial linear co-polarized Raman response, which is expected from the ${\mathrm{A}_{1}}^{\prime}$ mode. Accordingly, the mode observed at $\SI{247.9}{\per\centi\meter}$ could be potentially assigned to the ${\mathrm{E}}^{\prime}$. A splitting of the ${\mathrm{A}_{1}}^{\prime}$ and ${\mathrm{E}}^{\prime}$ modes under uniaxial strain has previously been predicted in the literature\cite{sahin_anomalous_2013}. Furthermore, the linewidth of the peak at $\SI{251.1}{\per\centi\meter}$ is narrower than those of the other observed Raman modes, consistent with a first-order Raman scattering process. In this case, only phonons from a small region around the $\Gamma$ point contribute to the scattering process, where the phonon branches are only weakly dispersive. In contrast, resonant second-order Raman scattering processes probe substantially larger regions of the the phonon dispersion, resulting in broader spectral features.

The spectral region usually assigned to the 2LA(M) mode complex has previously been analyzed in detail and assigned to a superposition of the 2LA(M), ZO(K)(A2''), LO(K)(E'), ZO(K)(A1') and ZO(M)(A2'') modes \cite{de_luca_new_2020}. In our lineshape analysis, this spectral region is described by three Lorentzian peaks. An even closer agreement with the literature could likely be achieved by introducing additional Lorentzian peaks to the lineshape analysis. However, this would reduce the robustness and stability of the fitting procedure and was therefore avoided.

The interpretation presented in the main text suggests that phonons near the $Q$ point contribute to the third-order Raman scattering processes in our resonance Raman measurements through a double resonance by coupling of the energetically close bright exciton $\mathrm{X_{KK}}$ and momentum-dark exciton $\mathrm{X_d}$. If this interpretation is correct, $Q$-point phonos may also contribute to Raman spectra obtained under resonant excitation close to the bright exciton $\mathrm{X_{KK}}$. In contrast to phonon modes at the $M$ and $K$ points, $Q$-point phonons can benefit from a double-resonance enhancement, potentially leading to a significantly larger scattering amplitudes. Consequently, the set of experimentally observable Raman modes may differ between resonant and non-resonant excitation conditions, despite the comparatively low phonon density of states near the $Q$ point.

\section{Determination of resonance profiles}

\begin{figure*}[h]
    \includegraphics[]{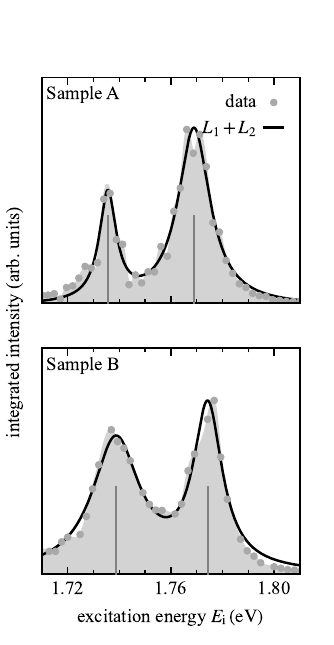}
    \caption{\label{fig:Figure_S2}
    Resonance profile of the $\AEphonon$ peak intensity for sample A (top) and sample B (bottom). Grey data points are shown with a guide to the eye. The black curve shows the total fit using the sum of two Lorentzians $L_{\mathrm{1}}+L_{\mathrm{2}}$. The grey vertical line marks the central energy of the Lorentzian.}
\end{figure*}

The resonance profile of the integrated $\AEphonon$ peak intensity shows a two-peak structure. To determine the positions and separation of the two peaks, the resonance profile is fitted with the sum of two Lorentzians $L_1(E_\mathrm{i})+L_2(E_\mathrm{i})$ where each Lorentzian $L_\mathrm{i}$ is defined by Eq. \ref{eq:lorentzian}. The fit results are shown in Fig. \ref{fig:Figure_S2} and the corresponding fit parameters are listed in the Tab \ref{tab:lorentz_fitparameter}. The deduced energy separation $\Delta E_0$ between the center energies of the two Lorentzian peaks is $\SI{33.4\pm0.4}{\milli\electronvolt}$ for Sample A and $\SI{35.6\pm0.5}{\milli\electronvolt}$ for Sample B, respectively. Both values exceed the simultaneously experimentally determined energy of the $\AEphonon$ phonon mode of $\SI{31.1}{\milli\electronvolt}$.

\begin{equation}
\label{eq:lorentzian}
    L(E_\mathrm{i})=\frac{A\mathit{\Gamma}}{2\pi}\frac{1}{\left(E_\mathrm{i}-E_0\right)^2+\left(\frac{\mathit{\Gamma}}{2}\right)^2}
\end{equation}

\begin{table}[h]
\caption{\label{tab:lorentz_fitparameter}Fit parameters obtained from line-shape analysis using the sum of two Lorentzian functions $L_1(E_\mathrm{i})+L_2(E_\mathrm{i})$ to reproduce the resonance profiles. The maximum of each resonance profile was normalized to unity prior to the fitting procedure. Consequently, the parameter $A$ does not represent the absolute peak intensities and is only meaningful with respect to the normalized data.}
\begin{ruledtabular}
\begin{tabular}{ccccc}
Parameter& \multicolumn{2}{c}{Sample A}& \multicolumn{2}{c}{Sample B}\\
& $L_\mathrm{1}$ & $L_\mathrm{2}$ &  $L_\mathrm{1}$ &  $L_\mathrm{2}$\\
\hline
$A (\mathrm{arb. units})$ & $\SI{0.778\pm0.079e-2}{}$ & $\SI{2.41\pm0.136e-2}{}$ & $\SI{2.673\pm0.163e-2}{}$ & $\SI{1.982\pm0.106e-2}{}$\\
$E_{0} (\SI{}{\milli\electronvolt})$ & $1735.6\pm0.3$ & $1769.0\pm0.3$ & $1738.7\pm0.4$ & $1774.3\pm0.3$ \\
$\mathit{\Gamma} (\SI{}{\milli\electronvolt})$ & $8.0\pm1.0$ & $15.1\pm0.9$ & $22.2\pm1.5$ & $13.5\pm0.9$\\
$\Delta E_0$ (\SI{}{\milli\electronvolt}) & \multicolumn{2}{c}{$33.4\pm0.4$} & \multicolumn{2}{c}{$35.6\pm0.5$}
\end{tabular}
\end{ruledtabular}
\end{table}

\section{Dependence of the expected splitting between incoming and outgoing resonance on the exciton linewidth}

In the main text, we demonstrated that the experimentally observed resonance peak separation cannot be reproduced by the first-order Raman scattering model. In a first order Raman process the double-peak structure is given by incoming and outgoing resonance conditions with a real excitonic intermediate state. The model predicts that the incoming and an outgoing resonance are separated by the phonon energy $E_\mathrm{ph}=E_{\scriptstyle{\mathrm{A}_{1}}^{\prime}/\mathrm{E}^{\prime}}$ as illustrated schematically in Fig. 2(a). The experimentally observed peak separation exceeds this phonon energy. This simplified picture neglects the finite exciton linewidth $\mathit{\Gamma}_\mathrm{X}$. For simplicity, this effect was omitted in the main text and is discussed here.

\begin{figure*}[h]
    \includegraphics[]{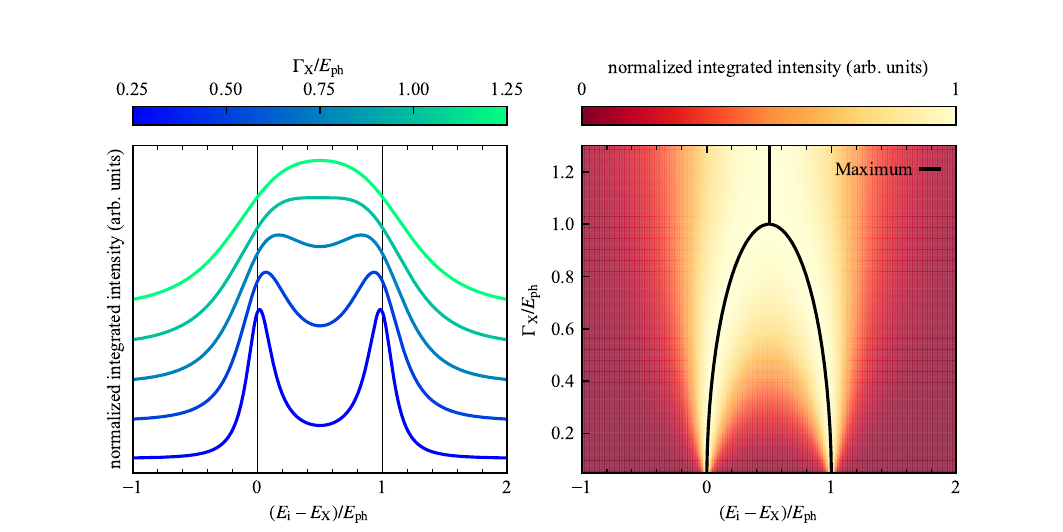}
    \caption{\label{fig:gamma_study1} Scattering probability of the first-order Raman model $P_{\mathrm{1^{st}}}$ as a function of the exciton linewidth $\mathit{\Gamma}_\mathrm{X}$ neglecting the correction function $\mathcal{F}$. Left panel: the calculated resonance profiles are normalized and vertically offset for clarity and the color code shows the exciton linewidth $\mathit{\Gamma}_\mathrm{X}$ as a function of the phonon energy $E_\mathrm{ph}$. Right panel: false color representation of the resonance profiles showing the exciton linewidth $\mathit{\Gamma}_\mathrm{X}$ dependence of the maxima highlighted by black lines. The scattering probability for each ratio is normalized to 1.}
\end{figure*}

\begin{figure*}[h]
    \includegraphics[]{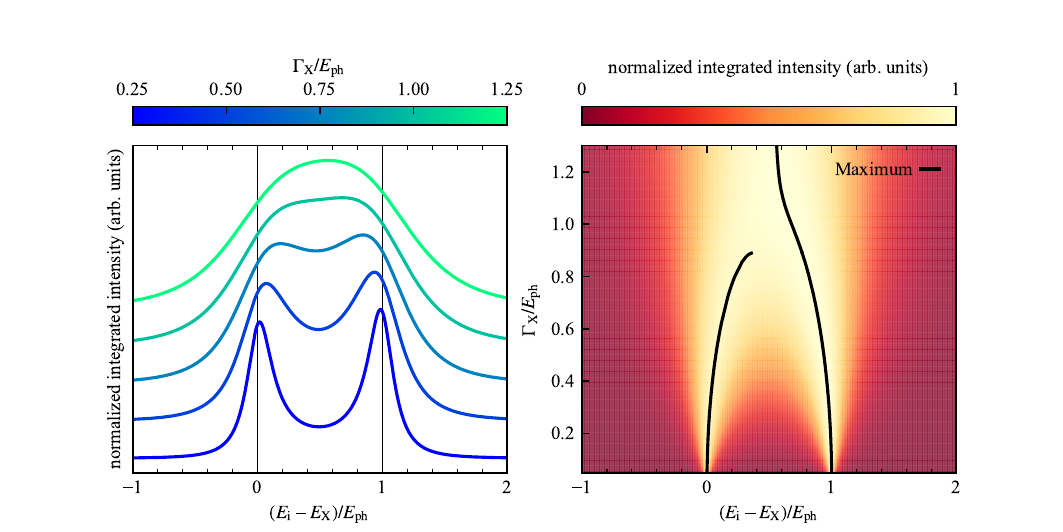}
    \caption{\label{fig:gamma_study2} Scattering probability of the first-order Raman scattering model $P_{\mathrm{1^{st}}}$ as a function of the exciton linewidth $\mathit{\Gamma}_\mathrm{X}$ taking the correction function $\mathcal{F}$ into account. \ref{fig:gamma_study1}}.
\end{figure*}

In Fig. \ref{fig:gamma_study1}, the calculated scattering probability according to the first-order Raman scattering model $P_{\mathrm{1^{st}}}$ neglecting the correction function $\mathcal{F}$, is shown as a function of the a ratio  $\mathit{\Gamma}_\mathrm{X}/E_\mathrm{ph}$ showing the dependence on the exciton linwidth $\mathit{\Gamma}_\mathrm{X}$. In Fig. \ref{fig:gamma_study2} the corresponding results are shown with the correction function $\mathcal{F}$ taken into account.

The finite linewidth $\mathit{\Gamma}_\mathrm{X}>0$ of the exciton $\mathrm{X}_\mathrm{KK}$ affects the scattering probability of the first-order Raman scattering model $P_{\mathrm{1^{st}}}$ by broadening both resonances resulting in an increased overlap between one resonance peak and the tail of the other resonance. Consequently, the maxima of the two resonance peaks (black lines in the false-color plot) shift towards each other and the minimum becomes shallower. When the linewidth $\mathit{\Gamma}_\mathrm{X}$ is equal to or larger than the phonon energy $E_\mathrm{ph}$, the two maxima merge into a single one. In the case where the correction function $\mathcal{F}$ is neglected, the position of the maxima can be calculated analytically, yielding the peak separation

\begin{equation}
    \label{eq:deltaE}
    \Delta E=\sqrt{E_\mathrm{ph}^2-\mathit{\Gamma}_\mathrm{X}^2}
\end{equation}

Using the phonon energy $E_\mathrm{ph}=\SI{31.1}{\milli\electronvolt}$ and the exciton linewidth $\mathit{\Gamma}_\mathrm{X}=\SI{14}{\milli\electronvolt}$ estimated from the Lorentzian fit presented in Tab. \ref{tab:lorentz_fitparameter} the ratio yields $\mathit{\Gamma}_\mathrm{X}/E_\mathrm{ph}=0.45$. According to Eq. \ref{eq:deltaE}, the peak splitting of the resonance profile is thereby reduced to $\SI{27.8}{\milli\electronvolt}$, corresponding to a reduction of approximately $10\%$. Consequently, the discrepancy between the experimentally observed peak splitting in the resonance profiles and the prediction of the first-order Raman scattering model becomes even larger.

The qualitative behavior persists, when the correction function $\mathcal{F}$ is included, as shown in Fig. \ref{fig:gamma_study2}. The calculated resonance profiles become asymmetric, which causes the lower-energy resonance peak (incoming resonance) to disappear rather than merge with the higher energy resonance peak (outgoing resonance). 

\section{Additional 3rd order scattering channel}

\begin{figure*}[h]
    \includegraphics[]{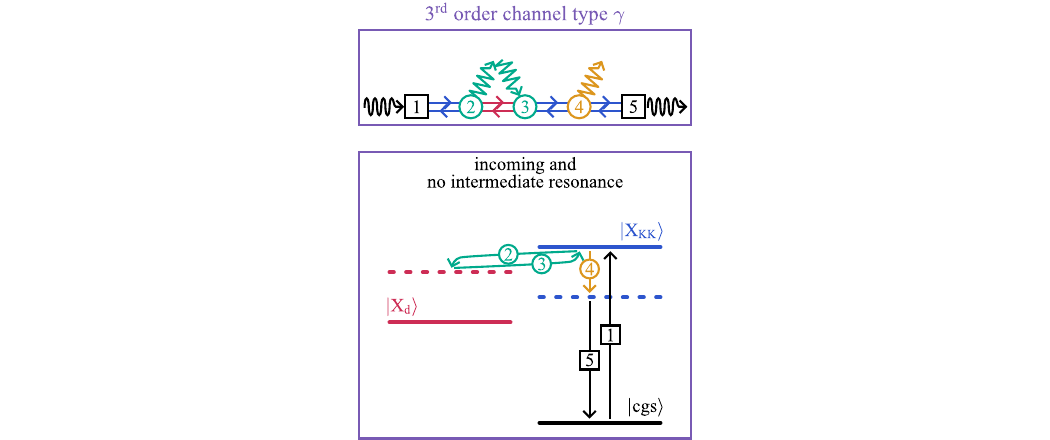}
    \caption{\label{fig:Figure_S1}
    3\textsuperscript{rd} order Raman scattering channel type $\gamma$ via bright and momentum dark excitons in analogy to Fig. 2 in the emain text. 
    3\textsuperscript{rd} order channel type $\gamma$: In contrast to the channel type $\beta$ the two finite-momentum phonon $\kappa$ (green) scattering events precede the $\AEphonon$ (orange) phonon emission. For the same detuning as in Fig. 2 no resonance with the momentum-dark exciton state $\left|\mathrm{X_{d}}\right>$ (red) occurs but the bright exciton $\left|\mathrm{X_{KK}}\right>$ (blue) serves twice as a resonant intermediate state.}
\end{figure*}

In the main text, three 3\textsuperscript{rd} order Raman scattering channels are discussed. However, only two of them (type $\alpha$ and type $\beta$) are depicted in Fig. 2. The missing 3\textsuperscript{rd} order channel type $\gamma$ is shown in Fig \ref{fig:Figure_S1}. 
Following photon absorption, the population the bright exciton state $\left|\mathrm{X_{KK}}\right>$ is populated. Subsequently, two finite-momentum phonon $\kappa$ scattering events first cause scattering to the momentum-dark exction state $\left|\mathrm{X_d}\right>$ and then repopulate the bright exciton state $\left|\mathrm{X_{KK}}\right>$. Finally, an $\AEphonon$ phonon is emitted from the bright exciton state $\left|\mathrm{X_{KK}}\right>$, followed by the emission of the scattered exciton.

The depicted incoming resonance contains a second resonance, since the bright exciton state $\left|\mathrm{X_{KK}}\right>$ is populated twice with the same energy. In contrast, the momentum-dark exciton state $\left|\mathrm{X_{d}}\right>$ is only populated non-resonantly. The momentum-dark exction state $\left|\mathrm{X_{d}}\right>$ is depicted according to the detuning $\delta$ values obtained from the fit to the resonance profile using the scattering probability of the intervalley scattering model $P_{\mathrm{1^{st}+3^{rd}}}$ as presented in the main text.

\section{Fitparamters of the intervalley scattering model}

\begin{table}[h]
\caption{\label{tab:fitparameter}Parameters from fitting the scattering probability $P_{\mathrm{1^{st}+3^{rd}}}$ to the resonance profiles. Complex numbers are modeled using the polar representation $z=|z|\cdot e^{i 2\pi\varphi_z}$}
\begin{ruledtabular}
\begin{tabular}{ccc}
fitted parameter&Sample A& Sample B\\
\hline
$E_\mathrm{X_{KK}} (\SI{}{\milli\electronvolt})$ & $1735.9\pm0.7$ & $1741.7\pm0.5$ \\
$\mathit{\Gamma}_\mathrm{X_{KK}} (\SI{}{\milli\electronvolt})$ & $21.4\pm1.5$ & $21.1\pm1.4$ \\
$\mathit{\Gamma}_\mathrm{X_{d}} (\SI{}{\milli\electronvolt})$ & $14.9\pm1.8$ & $17.1\pm2.0$ \\
$\delta (\SI{}{\milli\electronvolt})$ & $44.9 \pm 0.7$ & $54.7 \pm 1.2$ \\
$|a| (\mathrm{arb. unit})$ & $\SI{1.03 \pm 0.10e-10}{}$ & $\SI{2.43 \pm 0.11e-10}{}$ \\
$|b| (1)$ & $0.28 \pm 0.20$ & $1.80 \pm 0.42$ \\
$\varphi_b (1)$ & $0.226\pm0.133$ & $0.156 \pm 0.036$ \\
$|c| (\SI{}{\milli\electronvolt^2})$ & $186\pm28$ & $274\pm56$ \\
$\varphi_c (1)$ & $0.041\pm0.029$ & $0.055\pm0.018$ \\
\hline
fixed parameter&Sample A& Sample B\\
\hline
$E_{\scriptstyle{\mathrm{A}_{1}}^{\prime}/\mathrm{E}^{\prime}} (\SI{}{\milli\electronvolt})$ & $31.1$& $31.1$\\
$E_\mathrm{\kappa} (\SI{}{\milli\electronvolt})$ & $16.4$ & $16.4$ \\
$\mathrm{\varphi_a} (1)$ & $0$ & $0$ \\
\hline
calculated parameter&Sample A& Sample B\\
\hline
$E_\mathrm{X_{d}} (\SI{}{\milli\electronvolt})$ & $1690.9\pm1.0$ & $1687.0\pm1.3$ \\
$A_\mathrm{1^{st}} (\mathrm{arb. unit})$ & $\SI{1.06 \pm 0.21e-20}{}$ & $\SI{5.89 \pm 0.60e-20}{}$\\
$A_\mathrm{3^{rd}\alpha} (\mathrm{arb. unit})$ & $\SI{2.86 \pm 4.19e-29}{}$ & $\SI{1.43 \pm 0.89e-26}{}$\\
$A_\mathrm{3^{rd}\beta,\gamma} (\mathrm{arb. unit})$ & $\SI{3.64 \pm 1.34e-28}{}$ & $\SI{4.42 \pm 1.83e-27}{}$\\
\end{tabular}
\end{ruledtabular}
\end{table}

In the main text, only the energies $E$ and damping parameters $\Gamma$ obtained from fits of the the scattering probability $P_{\mathrm{1^{st}+3^{rd}}}$ of the intervalley scattering model are presented. Here, the amplitude scaling parameters $a$, $b$ and $c$ are additionally provided, which represent the interaction matrix elements as defined in Sec. \ref{sec:detailed formulas}. Tab. \ref{tab:fitparameter} lists the complete set of obtained parameters, including fixed parameters and quantities deduced from the fitted parameters.

The energy of the finite-momentum phonon $E_\mathrm{\kappa}$ was initially treated as a free parameter. However, this resulted in a strong correlation between $E_\mathrm{\kappa}$ and the detuning $\delta$. To avoid this ambiguity, $E_\mathrm{\kappa}$ was fixed at the energy of the prominent Raman peak discussed in Sec. \ref{sec:MFR} that is assigned to an acoustic phonon mode. The observed correlation indicates that choosing a finite-momentum phonon with a different energy would primarily lead to a corresponding shift in the fitted value of $\delta$.

The phase of $a$ is set to zero because the scattering probability, including the interference effects between the scattering channels, depends only on their relative phases. Consequently, one phase can be chosen freely without affecting the fit results.

The parameter $a$ determines the overall intensity scaling of the Raman scattering signal. While it is listed here for the sake of completeness, its value and order of magnitude depend on experimental setup-related factors and should therefore be interpreted with caution.

The amplitude scaling parameters provide quantitative information about the exciton-phonon coupling matrix elements. The absolute value of the matrix element $\left|\left<\mathrm{X_{KK}}\middle|H_\mathrm{ex.phon.}\middle|\mathrm{X_d}\right>\right|=\sqrt{\left|c\right|}$ is $\SI{13.6\pm 1.1}{\milli\electronvolt}$ for Sample A and $\SI{16.5\pm 1.7}{\milli\electronvolt}$ for Sample B. This indicates that the intervalley coupling enabled by finite-momentum phonon scattering is on the order of the finite-momentum phonon energy. The ratio of the matrix elements $\frac{\left|\left<\mathrm{X_{d}}\middle|H_\mathrm{ex.phon.}\middle|\mathrm{X_d}\right>\right|}{\left|\left<\mathrm{X_{KK}}\middle|H_\mathrm{ex.phon.}\middle|\mathrm{X_{KK}}\right>\right|}=\left|b\right|$ is $0.28\pm0.20$ for Sample A and $1.80\pm0.42$ for Sample B. These values indicate that the coupling of the optical phonon mode to the different excitonic states is sample dependent.






%

\end{document}